\documentclass{aa}  

\usepackage{graphicx}
\usepackage{txfonts}
\usepackage{natbib}
\bibpunct{(}{)}{;}{a}{}{,} 

\usepackage{hyperref}
\usepackage{placeins}
\hypersetup{
        colorlinks=true,
        breaklinks=true,
        citecolor=blue,
        allcolors=blue
}

\makeatletter
\renewcommand*\aa@pageof{, page \thepage{} of \pageref*{LastPage}}
\makeatother

\newcommand{\name}[1]{\textsc{#1}}

\begin{document}

   \title{Accretion onto supermassive and intermediate-mass black holes \\ in cosmological simulations}
   \titlerunning{Accretion in cosmological simulations}

   \author{R.~Weinberger
          \inst{1},
          A.~Bhowmick
          \inst{2},
          L.~Blecha
          \inst{3},
          G.~Bryan
          \inst{4},
          J.~Buchner
          \inst{5},
          L.~Hernquist
          \inst{6},
          J.~Hlavacek-Larrondo
          \inst{7},
          V.~Springel
          \inst{8}
          }
  \authorrunning{R. Weinberger et al.}

   \institute{Leibniz Institute for Astrophysics Potsdam (AIP),
              An der Sternwarte 16, 14482 Potsdam, Germany\\
              \email{rweinberger@aip.de}
         \and
             Department of Astronomy, University of Virginia, Charlottesville, VA 22904, USA
         \and
             Department of Physics, University of Florida, Gainesville, FL 32611, USA 
         \and
             Department of Astronomy, Columbia University, 550 West 120th Street, New York, NY 10027, USA
         \and
             Max-Planck-Institute for Extraterrestrial Physics, Gie\ss enbachstraße 1, 85748 Garching, Germany
         \and 
             Center for Astrophysics $\left| \right.$ Harvard \& Smithsonian, 60 Garden Street, Cambridge, MA 02138, USA
         \and
             D\'epartement de Physique, Universit\'e de Montr\'eal, Succ. Centre-Ville, Montr\'eal, Qu\'ebec H3C 3J7, Canada
         \and
            Max Planck Institute for Astrophysics, Karl-Schwarzschild-Str 1, D-85741 Garching, Germany
             }

   \date{Received September 15, 1996; accepted March 16, 1997}

  \abstract{ 
   Accretion is the dominant contribution to the cosmic massive black hole (MBH) density in the Universe today. However, modelling accretion in cosmological simulations is challenging due to the dynamic range involved, as well as the theoretical uncertainties of the underlying mechanisms driving accretion from galactic to black hole horizon scales.
   We present a simple, flexible parametrisation for gas inflows onto MBHs aimed at managing this uncertainty in the context of large-volume cosmological simulations. This study has been carried out as part of the `Learning the Universe' collaboration,  with an aim to jointly infer the initial conditions and physical processes governing the evolution of the Universe, using a Bayesian forward-modelling approach. 
   To allow for this forward-modelling approach, we updated the prescription for accretion with a two-parameter free-fall based inflow estimate that allows for a radius-dependent inflow rate and added a simple model for unresolved accretion disks.
   We used uniform resolution cosmological hydrodynamical simulations and the IllustrisTNG framework to study the MBH population and its dependence on the introduced model parameters. 
   Once the parameters of the accretion formula were chosen, aimed at achieving a zero black hole mass density (BHMD) at a roughly similar redshift, the differences caused by  details in the accretion formula are moderate in the supermassive black hole (SMBH) regime, indicating that it is difficult to distinguish between accretion mechanisms based on luminous active galactic nuclei (AGNs) powered by SMBHs.
   Applying the same models to intermediate-mass black holes (IMBHs) at high redshifts, however, reveals significantly different accretion rates in high-redshift, moderate-luminosity AGNs, as well as different frequencies and mass distributions of IMBH mergers for the same black hole formation model. This difference in the early growth history will also likely  lead to an accretion model-dependent SMBH population in low-mass black hole formation scenarios.
   }


   \maketitle

\section{Introduction}

Massive black holes (MBHs) are among the most poorly understood objects in extragalactic astrophysics. While their existence can be established using stellar dynamics in the galactic centre \citep{Ghez2008, Genzel2010} or emission of the inner accretion disk \citep{EventHorizonTelescopeCollaboration2019}, their evolution is relatively poorly constrained \citep{Greene2020, Inayoshi2020, Harikane2023, Taylor2024, Matthee2024}.
Studies of luminous active galactic nuclei (AGNs) show that most of their overall mass growth is dominated by gas accretion in bright quasi-stellar object phases \citep{Soltan1982, Yu2002}, with the hard X-ray background dominated by the cosmic black hole accretion rate density (BHARD; \citealt{Ueda2003, Shankar2009}). Comparing its redshift dependence to the star formation rate density \citep{Merloni2004} hints at a likely co-evolution scenario for supermassive black holes (SMBHs) and their host galaxies, with the most massive MBHs assembling earlier than the lower mass ones \citep{Shankar2004}.

Mergers with other MBHs are a natural consequence of hierarchical structure formation and represent a second growth channel. While overall subdominant \citep{Ni2022}, this channel can contribute substantially to the mass growth of MBHs in specific evolutionary stages, assuming the inspiral time of a MBH binary from kpc scales is sufficiently rapid:
since the host galaxy merger rate is dictated by cosmological structure formation, this primarily happens in phases where gas accretion is suppressed such as during early growth \citep{Bhowmick2024} and late-time evolution in massive galaxies and galaxy clusters \citep{Weinberger2018}.

The masses of MBHs show correlations with the stellar content of their host galaxies \citep{Magorrian1998}, nuclear kinematics \citep{Ferrarese2000, Gebhardt2000, McConnell2013}, and other galactic properties \citep[e.g.][]{Graham2007, Martin-Navarro2018}. The question of how we can  interpret these correlations is still a matter of debate \citep{Kormendy2013}, with galaxy mergers generally being seen to play a central role in them. On the one hand, they cause MBH mergers, which, in turn, narrow the distribution of MBHs to stellar mass ratios \citep{Peng2007, Jahnke2011, Graham2023}. On the other hand, they trigger central starbursts and (possibly) quasar activity \citep{DiMatteo2005, Hopkins2006, Hopkins2008}, manifested as correlations between luminous AGNs and mergers \citep[]{Ellison2019, Marian2020}. Yet, even when taking the cosmological formation history into account, the interpretation of these correlations still depends on the assumed gas accretion rate from galactic to accretion disk scales, most notably its MBH mass dependence \citep{Angles-Alcazar2013}.

In the early Universe, high redshift quasars powered by rapidly accreting SMBHs are detected up to and exceeding $z=7$ \citep[see][and references therein]{Inayoshi2020, Fan2023}. These objects represent the most extreme formation and growth scenarios of MBHs, with hints of a significantly more abundant population of lower mass, high redshift AGN recently being found due to better observational capabilities \citep{Maiolino2023, Adamo2024, Greene2024, Kokorev2024, Kocevski2024}. While the mass estimates of the MBHs powering these sources are highly uncertain, the existence of these luminous AGNs seems to be a challenge to commonly assumed evolutionary pathways \citep{Akins2024, Durodola2024, Kovacs2024}. This concerns in particular the MBH-galaxy scaling relations of these objects, indicating that these MBHs are substantially over-massive, even when taking the substantial uncertainties in MBH and stellar mass measurements into account \citep{Pacucci2023, Natarajan2024}.

At the opposite extreme, we have MBHs that have never undergone any substantial growth period and remain at relatively similar masses until $z=0$ \citep[see][for a review]{Greene2020}. These intermediate-mass black holes (IMBHs) are  expected to be found in dwarf galaxies or globular clusters \citep{Mezcua2017}; however,  unambiguous evidence is still lacking \citep[see, however,][]{Haberle2024}. From a MBH evolution perspective, these `leftover' IMBHs are the slowest growing and thus cleanest remnants of the formation process of MBHs in the high redshift Universe, thus a promising avenue to distinguish between possible formation scenarios \citep[see][for a review]{Volonteri2021}. Throughout this paper, we use MBH as the general term encompassing both IMBHs with mass $M<10^6$~M$_\odot$ and SMBHs with $M\geq 10^6$~M$_\odot$.

Understanding accretion onto MBHs from a theoretical perspective is challenging not only due to the dynamic range from galactic to black hole (BH) scales, but also due to a variety of other physical processes that simultaneously happen in galactic centres. However, substantial progress has been made in the past few years simulating accretion flows from galactic to accretion disk scales, in large parts driven by advances in adaptive refinement \citep{Curtis2015, Angles-Alcazar2021} and improved algorithms to cover the necessary dynamic range.
Studies of quasars \citep{Curtis2016, Hopkins2024, Hopkins2024b, Lupi2024} as well as accretion flows onto elliptical galaxies \citep{Guo2023, Guo2024}, galaxy groups and clusters \citep{Gaspari2017}, and Milky-Way mass galaxies \citep{Emsellem2015, Ressler2018} have all demonstrated that it is possible to study specific environments of MBHs. These approaches have provided substantial insights into the nature of accretion flows. Still, gas inflows in these different environments seem to be dominated by different aspects (cooling time, gravitational torques, gas supply, magnetic fields, etc.) and a unified picture has not yet emerged.

In cosmological simulations of galaxy formation, MBHs have become an essential ingredient \citep[see][and references therein]{DiMatteo2023}. The AGN feedback invoked in the simulation models proves critical for reproducing the high mass end of the galaxy population \citep{Sijacki2007, Weinberger2017} and determining galaxy cluster properties \citep{Barnes2018}. It even affects cosmological structure formation per se \citep{Springel2018}. Solving for the cosmological evolution of halos, their host galaxies with gas cooling, star formation, and gas inflows onto the central MBHs over cosmic time yields a wealth of predictions about MBHs and their host galaxies which can be compared to observations \citep{Habouzit2019, Terrazas2020, Li2020}. This makes MBHs a powerful predictive tool, provided the employed model is realistic and accurate.

One notable incompleteness of many current state-of-the-art galaxy formation models are IMBHs and the ability to represent formation channels that result in lower-mass `seed' BHs that could range from $10^2-10^4$~M$_\odot$. This is partially due to computational limitations, where the mass of a resolution element is often larger than the respective formation mass. However, the MBH model might itself be an obstacle due to the employment of the Bondi-Hoyle formula \citep{Bondi1944, Bondi1952} as the accretion estimate \citep[see however][who use a different accretion model]{Dave2019}. The quadratic scaling of the accretion rate with BH mass leads to a suppression of accretion for low-mass IMBHs, especially when employed in combination with an effective equation of state model for the interstellar medium (ISM) that provides a floor to the sound speed \citep{Booth2009}. This leads to a growth history that is too delayed to explain high-redshift luminous AGNs, unless massive seeds are invoked. While IMBHs can still grow significantly via mergers in permissive seeding models that predict frequent seed formation \citep{Bhowmick2024b}, an alternative solution to this problem cold be that the Bondi-Hoyle formula is simply inadequate for this early growth regime. Since neither our theoretical understanding, nor our observational capabilities are able to rule out either scenario directly, we need to model both in cosmological simulations to explore indirect effects.

This work is part of the `Learning the Universe' collaboration\footnote{https://www.learning-the-universe.org} dedicated to understanding the extragalactic Universe by jointly inferring the initial conditions and physical laws that govern its subsequent evolution. As part of the effort to build the predictive galaxy formation model needed for this endeavour, we present a new, flexible parametrisation for gas inflows and MBH growth via accretion in large-volume cosmological simulations. In the first step, the gas properties are measured at a fixed, galactic scale, assumes a scaling of the mass inflow rate with radius, and feeds a simple prescription for accretion disk evolution, which ultimately grows the MBH. The paper is structured as follows. Section~\ref{sec:model} describes our numerical setup and models, Section~\ref{sec:results} presents the results. We discuss our findings in Section~\ref{sec:discussion} and present our conclusions in Section~\ref{sec:conclusions}.

\section{Model}
\label{sec:model}

\begin{figure*}
    \includegraphics[]{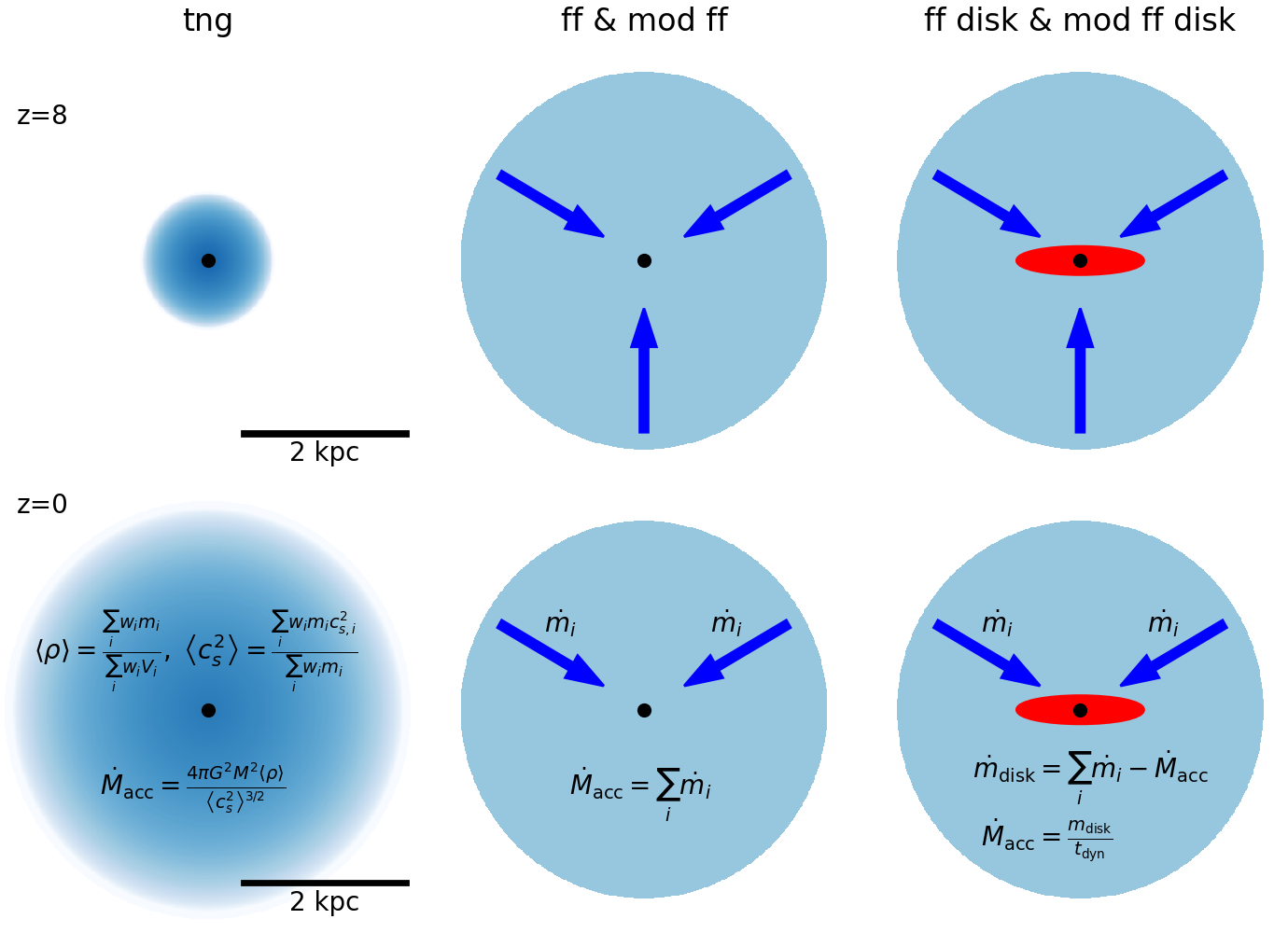}
    \caption{Schematic of the different accretion estimates used in this work. Left: Reference method of the \texttt{tng} model, in the centre the \texttt{ff} and \texttt{mod ff} models and on the right the models with accretion disk (\texttt{ff disk} and \texttt{mod ff disk}). Top and bottom rows show the kernels at $z=8$ and $z=0$, respectively, both at fixed proper distances. The change in kernel size in \texttt{tng} is the median from the respective L50n768 simulation, while the median kernel size in the other models is constant.}
    \label{fig:schematic}
\end{figure*}

\subsection{Initial conditions}

We ran uniform resolution cosmological simulations starting at $z=127$, created by adopting the \citet{PlanckCollaboration2016} $\Lambda$ cold dark matter cosmological parameters ($\Omega_\textrm{m} = 0.3089$, $\Omega_\textrm{b} = 0.0486$, $\Omega_\Lambda = 0.6911$, $H_0 = 67.74$~km~s$^{-1}$~Mpc$^{-1} = 100\,h$~km~s$^{-1}$~Mpc$^{-1}$, $\sigma_8 = 0.8159,$ and $n_s=0.9667$). We also used the same linear theory power spectrum as the \name{IllustrisTNG} simulations. The initial displacement is calculated using the \name{NgenIC} code in the version originally developed for \name{Gadget4} \citep{Springel2021} using second-order Lagrangian perturbation theory and applying a variance suppression technique \citep{Angulo2016} to enhance the representativeness of the simulated volume. The gas was initialised from the same density and velocity field, splitting the matter into an equal number of gas cells and dark matter particles and distributing the mass according to the cosmic baryon and dark matter fractions, respectively. A uniform magnetic field was initialised with $B = 10^{-14}$~co-moving Gauss. 

In the following, we present cosmological simulations with two different volumes. Specifially, to understand the overall evolution over cosmic time, we evolved boxes of sidelength $50\,h^{-1}$~Mpc (L50) to a redshift of $z=0.$ We probed the structure formation of relatively massive halos, but at comparably low mass resolution. To study the high-redshift, early growth of IMBHs, we evolved smaller volumes with sidelength of $12.5\,h^{-1}$~Mpc (L12.5) to redshift $z=5$. These simulations adopted lower MBH seed masses and were used to study the IMBH growth at high redshifts. All simulations were run with $2\times 192^3$, $2\times 384^3$ and $2\times 768^3$ initial resolution elements, covering  a dynamic range of $8^4 = 4096$ in mass resolution overall and an identical mass resolution of the lowest resolution (L12.5) and the highest resolution (L50) simulation. Unless stated explicitly, we show the highest resolution version of the respective simulation box.

   \begin{table}
      \caption[]{Simulation initial conditions.}
         \label{tab:ics}
     
         \begin{tabular}{l c c c}
         \hline
         \noalign{\smallskip}
              Name & $\epsilon$  & $m_\mathrm{dm}$  & $m_\mathrm{target}$ \\
                   & [$h^{-1}$~kpc] & [$10^6$~$h^{-1}$~M$_\odot$] & [$10^5$~$h^{-1}$~M$_\odot$] \\
              \noalign{\smallskip}
              \hline
              \noalign{\smallskip}
              L12.5n768 & $0.4$ & $0.31$ & $0.58$\\
              L12.5n384 & $0.8$ & $2.49$ & $4.65$\\
              L12.5n192 & $1.6$ & $19.9$ & $37.2$\\
              L50n768 & $1.6$ & $19.9$ & $37.2$\\
              L50n384 & $3.2$ & $159$ & $298$\\
              L50n192 & $6.4$ & $1276$ & $2382$\\
              \noalign{\smallskip}
              \hline
         \end{tabular}
        \tablefoot{
            The name encodes the box side length in h$^{-1}$~Mpc following L and the number of resolution elements per dimension following n. At startup, the matter particles are split into an equal number of dark matter particles of mass, $m_\mathrm{dm}$, and gas cells of mass, $m_\mathrm{target}$, according to the chosen cosmic baryon fraction. All lengths are measured in co-moving coordinates.
        }
   \end{table}

\subsection{Reference: The  IllustrisTNG model}

The simulations were carried out using the \name{Arepo} code \citep{Springel2010, Pakmor2011, Pakmor2016, Weinberger2020} solving the  magneto-hydrodynamics (MHD) equations using the finite-volume technique on a moving mesh and the Poisson equation using a tree-particle-mesh method. An eight-wave Powell cleaning scheme was applied to ensure approximately divergence-free magnetic fields \citep{Pakmor2013}.

Our main aim is to develop and test a new model for accretion onto MBHs in cosmological environments. We used the IllustrisTNG galaxy formation model as a well-studied framework with comprehensive analysis of the behaviour of its simulated MBH population in the literature \citep[e.g.][among many others]{Bhowmick2020, Terrazas2020, Li2020, Truong2021, Habouzit2021, Habouzit2022, Habouzit2022b}. 
In the following, we briefly summarise the IllustrisTNG model. For a more detailed description, we refer to the IllustrisTNG method papers \citep{Weinberger2017, Pillepich2018}.

We included radiative cooling down to $10^4$~K using primordial \citep{Katz1996} and metal-line cooling \citep{Wiersma2009}, taking into account a time dependent spatially homogeneous UV background \citep{Faucher-Giguere2009}\footnote{Note that revised backgrounds exist \citep{Puchwein2019, Faucher-Giguere2020}. However, to be consistent with IllustrisTNG we use the original model.} and adding corrections for self-shielding \citep{Rahmati2013}. Additionally, the radiation field of nearby AGN is taken into account. The implementation details are described in \citet{Vogelsberger2013}.

The star formation was modelled following \cite{Springel2003, Vogelsberger2013}, with dense gas above a threshold density of $n=0.13$~cm$^{-3}$ established as the star-forming limit. We used an effective equation of state (or in a hotter state), which allowed us to model the unresolved ISM. Star particles form in a stochastic manner from star forming gas and are evolved assuming a single stellar population with \citet{Chabrier2003} initial mass function, taking into account chemical enrichment and mass return from asymptotic giant branch stars, core-collapse and type Ia supernovae \citep[with the TNG model updates as described in][]{Pillepich2018}. Stellar feedback is parametrised via massive, enriched galactic wind particles that are temporarily decoupled from the gas dynamics \citep{Springel2003, Vogelsberger2013} using the scaling of the IllustrisTNG simulation \citep{Pillepich2018}. 

The MBH modelling follows a relatively simple, numerically robust approach for seeding, dynamics, and mergers \citep{Springel2005, Sijacki2007, DiMatteo2008}.
Specifically, in every halo with a mass exceeding $5\times10^{10} h^{-1}$~M$_\odot$ that does not yet contain one, a MBH is seeded converting the most bound gas cell into the BH particle with seed mass $M = 8\times10^{5} h^{-1}$~M$_\odot$. This BH particle is kept at the potential minimum of the halo throughout the simulation, with MBH mergers being modelled instantaneously without explicitly following the dynamics of  BHs.

In the TNG model, gas accretion onto MBHs is modelled using the Eddington-limited Bondi formula applied to a kernel-averaged estimate of the gas density, $\rho$, and sound speed, $c_s$, in the cells surrounding the BH particle (average denoted by $\left< \right>$).
This equations are expressed as
\begin{align}
    &\dot{M}_\mathrm{acc}  = \min(\dot{M}_\mathrm{Bondi}, \dot{M}_\mathrm{Edd}), \\
    &\dot{M}_\mathrm{Bondi} = \frac{4 \pi G^2 M^2 \left<\rho\right>}{\left<c_s^2\right>^{1.5}},\\
    &\dot{M}_\mathrm{Edd}  = \frac{4 \pi G M m_p}{\epsilon_r \sigma_T c},
\end{align}
where $G$ is the gravitational constant, $M$ the MBH mass, $m_p$ the proton mass, $\epsilon_r=0.2$ the radiative efficiency, $\sigma_T$ the Thompson cross-section and $c$ the speed of light. The corresponding BH mass growth is
\begin{align}
    \dot{M} = (1-\epsilon_r) \dot{M}_\mathrm{acc}.
\end{align}

The averaging scale, $d,$ is adjusted on the fly at every time step using a target weighted number of cells, as 
\begin{align}
    n_\mathrm{ngb, target} = \sum\limits_j \frac{m_j}{m_\mathrm{target}} w (r_j, d) = 48\pm 1,
\end{align}
with a given tolerance\footnote{Note that we keep this number fixed with resolution, which is a deviation from the IllustrisTNG simulation, where it is scaled by a factor of $2$ for each resolution step (factor of $8$ in target mass).} and where $r_j$ denotes the distance of the cell mesh-generating point and the MBH position and $m_\mathrm{target}$ the target gas mass (i.e. the gas mass resolution). The weighting kernel is defined as

\begin{align}
    w(r, d) = \frac{8}{\pi d^3} 
    \begin{cases}
        1 - 6\left(\frac{r}{d}\right)^2 + 6\left(\frac{r}{d}\right)^3 \qquad &\textrm{for} \, \frac{r}{d} < 0.5, \\
        2 \left( 1 - \frac{r}{d}\right)^3 \qquad &\textrm{for} \, 0.5 \leq \frac{r}{d} < 1, \\
        0 \qquad &\mathrm{for} \, \frac{r}{d} \geq 1.
    \end{cases}
\end{align}

Feedback from AGNs is modelled in two modes \citep{Weinberger2017}.
Highly accreting MBHs relative to the Eddington limit inject thermal energy continuously into their surroundings (using the same kernel)

\begin{align}
    \dot{E}_\mathrm{th} = \epsilon_{f,\mathrm{high}} \epsilon_r \dot{M}_\mathrm{acc} c^2.
\end{align}

For low-accretion-rate MBHs, a kinetic kick is injected in a pulsed fashion whenever enough energy is available.
This is expressed as
\begin{align}
    \dot{E}_\mathrm{kin} = \epsilon_{f,\mathrm{kin}} \dot{M}_\mathrm{acc} c^2.
\end{align}

We note that both feedback channels are mutually exclusive, with an Eddington ratio threshold of 

\begin{align}
    \chi = \min\left[ 0.002 \left(\frac{M}{10^8\,\mathrm{M}_\odot}\right)^2, 0.1 \right].
\end{align}

This mass-dependent threshold is a key feature of the model, with AGN and their host galaxies transitioning from being luminous and star forming with young stellar populations to being low-luminosity and quiescent due to the significantly more efficient kinetic feedback \citep{Weinberger2018}.

\subsection{A new model for gas accretion onto MBHs}

In this work, we explore a novel way to calculate the accretion rate onto MBHs, while keeping all other aspects of the simulation the same. We note that our motivation for introducing this model is threefold. We aim to:\ 

\begin{enumerate}
    \item establish a better connection between cosmological simulations and simulations of smaller-scale nuclear gas inflows. To this end, we defined the region of surrounding gas from which the inflow is calculated with respect to a fixed proper distance from the MBH \footnote{As opposed to a (time) variable distance determined by a fixed weighted number of enclosed neighbouring cells commonly used in galaxy formation simulations.},  allowing for a radius-dependent mass inflow rate.
    \item absorb physically plausible scalings in model parameters that can be varied and constrained when confronted with data, while keeping the approach as simple as possible. Notably, we did not aim to obtain a method that closely approximates a specific physical mechanism driving inflows. This is because we deemed it unlikely that such a model would be appropriate in all cases in such large-scale cosmological simulations that follow the halos of a wide range of masses across cosmic time.
    \item develop a model that is ready to be extended and used with future modelling techniques. This includes a model for unresolved accretion disks, algorithmic changes to the time-integration more in line with the employed fluid-dynamics solver, and compatibility with multi-fluid modelling approaches \citep{Weinberger2023}. 
\end{enumerate}

\subsubsection{Fixed inflow scale}
Unlike the SPH-like kernel estimate used in the IllustrisTNG model, we define the surroundings of the BH as a sphere with fixed proper radius irrespective of the number of cells in this sphere. We used a radius of $d = 375 \, h^{-1}$~pc for L12.5n768 and $d = 1.5\, h^{-1}$~kpc for L50n768, and larger values for the n384 and n192 simulations, by a factor of $2$
and $4,$ respectively. This is several times the smoothing length of collisionless particles at high redshift and only slightly smaller than the smoothing length at $z=0$ (see Table~\ref{tab:ics}). We used a top-hat kernel over all cells in a sphere with radius, $d$, to estimate physical quantities. If there are no cells in this sphere, we increased $d$ to include at least one. This happens only for $1-2\%$ of BHs at high redshift. Due to a fixed target mass of cells, these are the cases where the gas density and, thus, the accretion rate is low and not significant for the overall growth. Starting at redshift $1$, $d$ needs to be increased more frequently, reaching up to several tens of percent at $z=0$, likely due to the evacuation of the central region by kinetic AGN feedback. An illustration of the different kernels is shown in Figure~\ref{fig:schematic}.

\subsubsection{Timescale of gas inflow}
The primary aim of our accretion estimate is to divide the available gas mass within a fixed aperture by the characteristic timescale of accretion from these scales. The question of which specific process sets this timescale at galactic scales remains open. Different processes might be relevant in different regimes; while \citet{Hopkins2010} investigated a post-merger scenario and argue that the transport of angular momentum via torques is the driving factor, \citet{Gaspari2013} investigated an efficiently cooling halo centre and argue that gas fuelling is dominated by cold clouds that chaotically fall in and lose their angular momentum via collisions (i.e. hydrodynamically). Studies of AGN feedback-regulated isolated galaxy clusters show that in this regime, the halo cooling rate is limiting the accretion \citep[e.g.][]{Meece2017, Ehlert2023}. A recent work by \citet{Hopkins2024} indicated that magnetic fields are crucial for accretion in quasar environments on sub-pc scales (neglecting, however, the effect of AGN quasar feedback in the simulation). 

For our model, we focus on the regimes most relevant for mass growth, namely, gas-rich environments with short cooling times. The Bondi accretion formula describes a spherically symmetric non-radiative accretion flow \citep{Bondi1952}. This is inadequate especially in a regime where radiative cooling is dominating the dynamics \citep{Gaspari2013}, as well as in regimes where the ISM treatment in the form of an effective equation of state interacts with the BH accretion by providing an artificial lower limit to the sound speed and upper limit to the density at fixed pressure. Previous works have pointed this out and applied corrections in terms of a density dependent boost factor \citep{Booth2009}. In this work, we dropped the use of the Bondi formula entirely and start from a different approach. We assumed that in the dominant growth phases:

\begin{enumerate}
    \item radial pressure gradients counteracting the inflow of gas can be neglected. In the case of a star-forming, multi-phase ISM with the different phases in hydrostatic equilibrium (as assumed for dense, star-forming gas in the simulation; \citealt{Springel2003}), this is a reasonable assumption for the colder phases that dominate the mass budget.
    \item angular momentum loss is efficient, leading to an inflow rate, $\dot{M}_\mathrm{acc}$, scaling with the dynamical or free-fall time, $t_\mathrm{ff}$ \citep[][their Eq. 69]{Hopkins2011},
as per    \begin{align}
        \dot{M}_\mathrm{acc} \sim \frac{\left| a\right| M_\mathrm{gas}}{t_\mathrm{ff}},
    \end{align}
    with $\left| a\right|$ is the fractional non-axisymmetric perturbation to the potential and $M_\mathrm{gas}$ the gas mass.
\end{enumerate}

Since we intend to use our accretion estimate in the context of large-volume samples and, consequently, in low-resolution simulations, we did not attempt to estimate $\left| a\right|$ or its scaling. Instead, we treated it as a constant. We consequently assume the accretion timescale to be proportional to the free-fall time onto the BH with mass, $M$,
\begin{align}
    t_\textrm{ff} = \left(\frac{d^3}{ G \, M }\right)^{1/2}.
\end{align}
This is similar to what was used in previous works on galaxy clusters \citep{Li2014, Ehlert2023} and in cosmological zoom simulations of galaxies \citep{Wellons2023}.
We note that we only use the MBH mass, not the enclosed mass of gas and stellar component in this timescale for simplicity and numerical robustness\footnote{Since the stellar mass scales with BH mass, its contribution can be to a degree absorbed in the normalisation factor later on. In the case of an unresolved accretion disk model, we used the combined mass of BH and accretion disk.}.

Furthermore, we did not explicitly model any mass outflow rates driven by unresolved stellar \citep{Hopkins2024, Partmann2025} or AGN feedback \citep{Ostriker2010, Choi2012, Cochrane2023,  Farcy2025}. While the normalisation factor absorbs effects linear to the enclosed mass, non-linear effects are not taken into account.

\subsubsection{Inflow rate estimate and its time integration}

We changed the time integration from a single estimate of the total inflow rate per MBH to an integration on every gas cell which is subsequently added up to the BH accretion rate. This allowed us to treat the inflow as a source term in the Euler equation and integrate it on its local time step. In the case where there are equal time steps for all cells in the surroundings of a MBH, both integration methods are equivalent.
The inflowing gas mass onto a MBH, $\Delta m$, is thus 
\begin{align}
    \Delta m = \sum\limits_i \Delta m_i,
\end{align}
where $\Delta m_i$ are the partial inflowing gas masses of individual active cells $i$ inside the top-hat kernel of the MBH.
For each cell, the accretion rate is integrated numerically on the local time step of the cell as a sink term in an operator-split fashion, thus
\begin{align}
    \Delta m_i = \int\limits_{t_n}^{t_{n+1}}  \dot m_i \, dt,
\end{align}
where
\begin{align}
    \dot{m_i} = \eta \, \frac{m_i}{t_\mathrm{ff}}.
\end{align}
We note that while the functional form appears similar to a star formation law, the free-fall time here is not the local free-fall time for the collapse of a self-gravitating cloud, but rather the free-fall time onto a point mass at distance $d$; i.e., not a local quantity.

Inspired by numerical studies of nuclear gas inflow \citep{Guo2023}, we additionally allow for a radius-dependent scaling of the inflow-rate, 

\begin{align}
    \eta = A \left(\frac{d}{R_s}\right)^\alpha
\end{align}
with $R_s$ being the Schwarzschild radius
\begin{align}
    R_s = \frac{2 G M}{c^2}.
\end{align}
Here, $\alpha$ and $A$ are the radial inflow scaling and the normalisation, respectively, and are treated as free parameters. This is slightly different from \citet{Guo2023} who suggest a scaling with Bondi-radius that would replace the speed of light $c$ by the sound speed at infinity $c_{s,\infty}$.  In either case, the scaling with a characteristic radius introduces a dependence on the MBH mass $M$ that can be controlled by varying $\alpha$. Thus, in practice the model has two free parameters: one for the normalisation and one for the MBH mass dependence.

In the following we present 2 inflow models to explore the impact of different mass dependences, one without inflow scaling $\alpha = 0$ as the simplest case, and one with the radial scaling of \citep{Guo2023}, $\alpha = -0.5$. In order to find the normalisation, different values for $A$ in $0.5$~dex steps were tested and the most similar result for the black hole mass density (BHMD) evolution at or slightly below the \texttt{tng} model chosen. We leave a detailed calibration to future work. The employed model parameters are listed in Table~\ref{tab:models}.

The inflowing gas $\Delta m$ is added to an unresolved accretion disk model (simulations \texttt{ff disk} and \texttt{mod ff disk}) or directly accreted (\texttt{ff} and \texttt{mod ff}). 

   \begin{table}
      \caption[]{Accretion model parameters}
         \label{tab:models}
         \begin{tabular}{l c c c}
         \hline
         \noalign{\smallskip}
              Name & $A$ & $\alpha$  & accretion disk model \\
              \noalign{\smallskip}
              \hline
              \noalign{\smallskip}
              \texttt{ff} & $0.001$ & $0$ & no\\
              \texttt{ff disk} & $0.001$ & $0$ & yes \\
              \texttt{mod ff} & $100$ & $-0.5$ & no\\
              \texttt{mod ff disk} & $100$ & $-0.5$ & yes\\
              \noalign{\smallskip}
              \hline
         \end{tabular}
   \end{table}

\subsubsection{Accretion disk model}

In two of the four new models presented in this work, we additionally use a model for unresolved accretion disks adopted from \citet{Wellons2023}. To do this, we add the inflowing mass $\Delta m$ to an accretion disk mass $m_\mathrm{disk}$, which itself fuels the accretion onto the BH,

\begin{align}
    \dot{M}_\mathrm{acc} &= \frac{m_\mathrm{disk}}{t_\mathrm{dyn}}, \\
    t_\mathrm{dyn} &= 42\,\mathrm{Myr} \left(1+\frac{M}{m_\mathrm{disk}} \right)^{0.4}.
\end{align}
 We note that $t_\mathrm{dyn}$ is the dynamical time of the accretion disk, which is different from the free-fall time from resolved scales. This model represents the \citet{Shakura1973} $\alpha$ disk model, however, is not intended to precisely reflect the complex accretion disk physics, but rather to broadly explore the impact of the use or the absence of an accretion disk model might have on the cosmic MBH population. While the most obvious will be a time-averaging, with fluctuations in the inflow rate on timescales below $t_\mathrm{dyn}$ being smoothed over in the accretion rate $\dot{M}_\mathrm{acc}$. Possible secondary effects on feedback remain less clear. It furthermore raises conceptual questions when applying an Eddington limit to the accretion rate, as we have done throughout this work.

\subsubsection{Eddington limited accretion}

One notable consequence of introducing an accretion disk model in global simulations is the question of where the Eddington limit has to be applied. Since the luminosity originates from the inner accretion disk and is proportional to the innermost accretion rate, it makes sense to limit this rate; i.e. the mass inflow rate in the unresolved accretion disk $\dot{M}_\mathrm{acc}$. The inflow rate estimate discussed earlier, however, is no longer directly linked to the resulting luminosity. This implies that there is no motivation to limit this inflow rate, $\dot{m}_i$, and we consequently refrain from doing so if the accretion disk model is active. A more detailed model might, however, account for the emitted radiation and its effect on the inflow rate, effectively describing unresolved feedback moderated accretion rates. In the spirit of keeping the model simple, and due to the need for a good theoretical understanding of how emitted radiation might modulate the inflow rate, we leave such developments for future work. In the simulations without an accretion disk model, we limit the inflow rate by rescaling $\dot{m}_i$, namely, rescaling the contribution of every cell to a given BH inflow rate by a constant factor in cases where the inflow rate estimate exceed the Eddington limit.

\subsection{AGN luminosity calculations}

Throughout this paper, we mostly show direct simulation output. Two notable exceptions are the AGN bolometric and X-ray luminosity.
To obtain bolometric luminosity, we follow \citet{Churazov2005, Hirschmann2014},
\begin{align}
    L_\mathrm{bol} = \epsilon_r \dot{M}_\mathrm{acc} c^2 \min \left(1, 10 \frac{\dot{M}_\textrm{acc}}{\dot{M}_\textrm{Edd}} \right).
\end{align}
From the bolometric luminosity, we obtain the $0.5-10$~keV X-ray luminosity by applying the bolometric corrections from \citet{Shen2020}.
This calculation is similar to other simulation analysis \citep[e.g.][]{Habouzit2022, Tremmel2024}.

\section{Results}
\label{sec:results}

In the following, we compare the new accretion prescription with the existing IllustrisTNG model (label \texttt{tng}). In this new model, we use a fixed aperture to determine the gas mass and the resulting free-fall time using only the BH mass, normalised with a constant $A=0.001$ assuming a constant mass inflow rate with $\alpha=0$ (label \texttt{ff}). As an alternative prescription, we use the same fixed aperture, but a radially dependent inflow rate scaling with $\alpha=-0.5$ and normalisation $A=100$ (label \texttt{mod ff}). For both these models, we run simulations with an additional accretion disk model to assess its impact (labels \texttt{ff disk} and \texttt{mod ff disk}, respectively). In the following, we study the impact of these models onto the SMBH population in the L50n768 simulations.

\subsection{SMBH evolution over cosmic time}

\begin{figure}
    \includegraphics[]{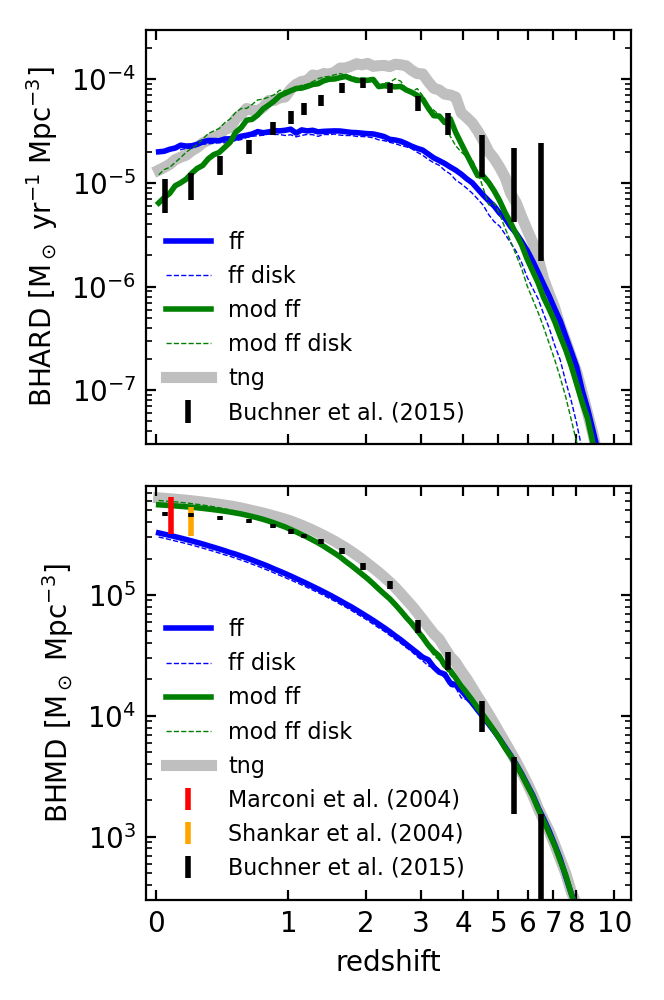}
    \caption{BHARD (top) and BHMD (bottom) vs redshift with different gas accretion models in the L50n768 simulations. The solid lines show the model without an unresolved accretion disk, the dashed respective simulations with a sub-grid accretion disk model. The grey line shows the respective version of the runs with the IllustrisTNG model, for reference. In red, yellow and black, we show measurements form \citet{Marconi2004, Shankar2004, Buchner2015}, respectively. All densities are measured in comoving coordinates.}
    \label{fig:bhard_z}
\end{figure}

\begin{figure*}
    \includegraphics[]{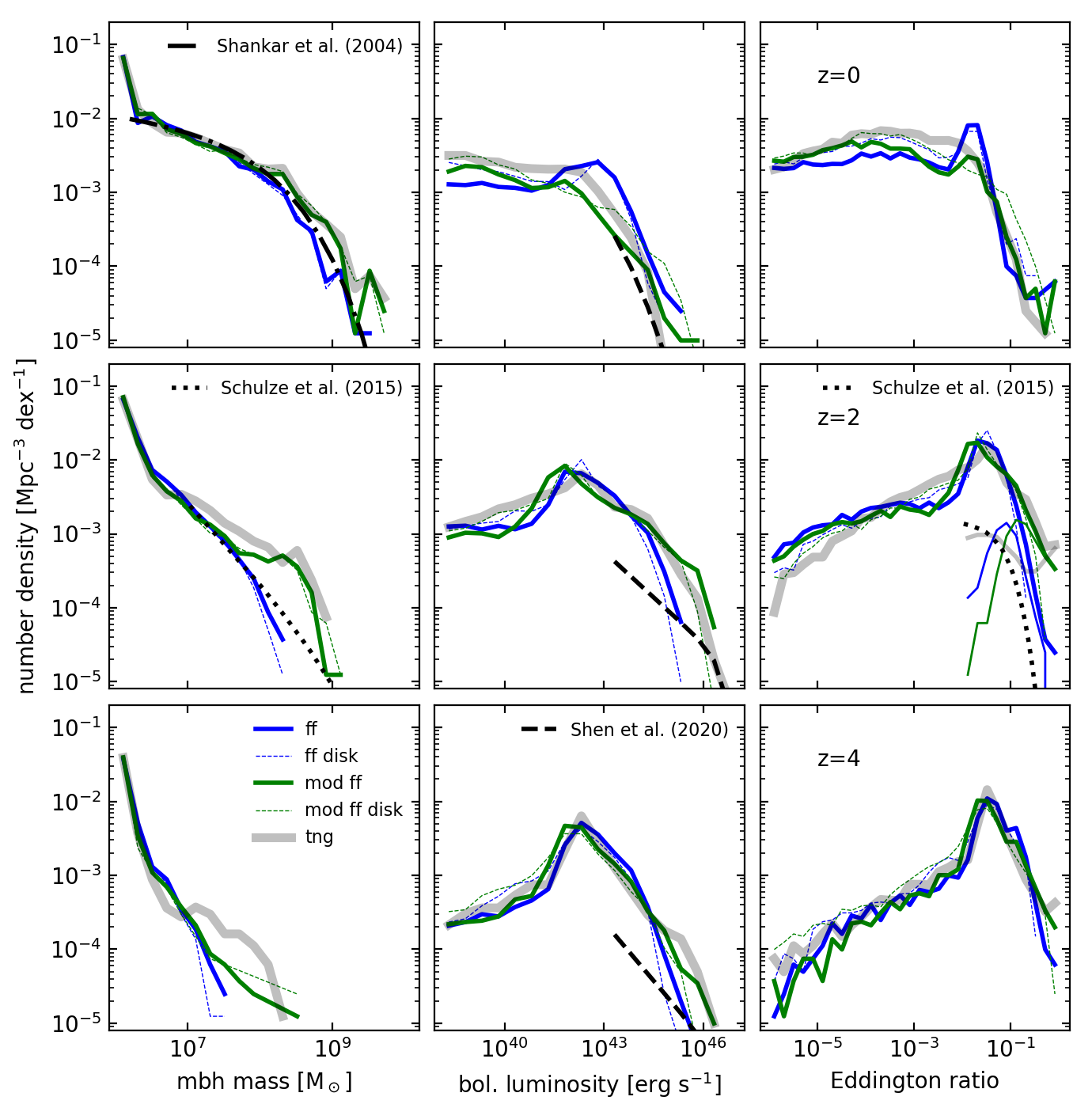}
    \caption{Black hole mass function (left), bolometric luminosity function (centre), and Eddington ratio distribution (right) at redshift $0$, $2$ and $4$ (top to bottom). The dashed lines show the respective models including an accretion disk model. Despite the very different nature of the accretion models, the resulting mass and luminosity functions are relatively similar. The second set of thinner lines in the Eddington ratio distribution are SMBHs with mass exceeding $10^7$~M$_\odot$ to be comparable to the literature values of \citet{Schulze2015}. Lines with accretion disk models are not shown for clarity.}
    \label{fig:mf_lf_xf}
\end{figure*}

\begin{figure*}
    \includegraphics[]{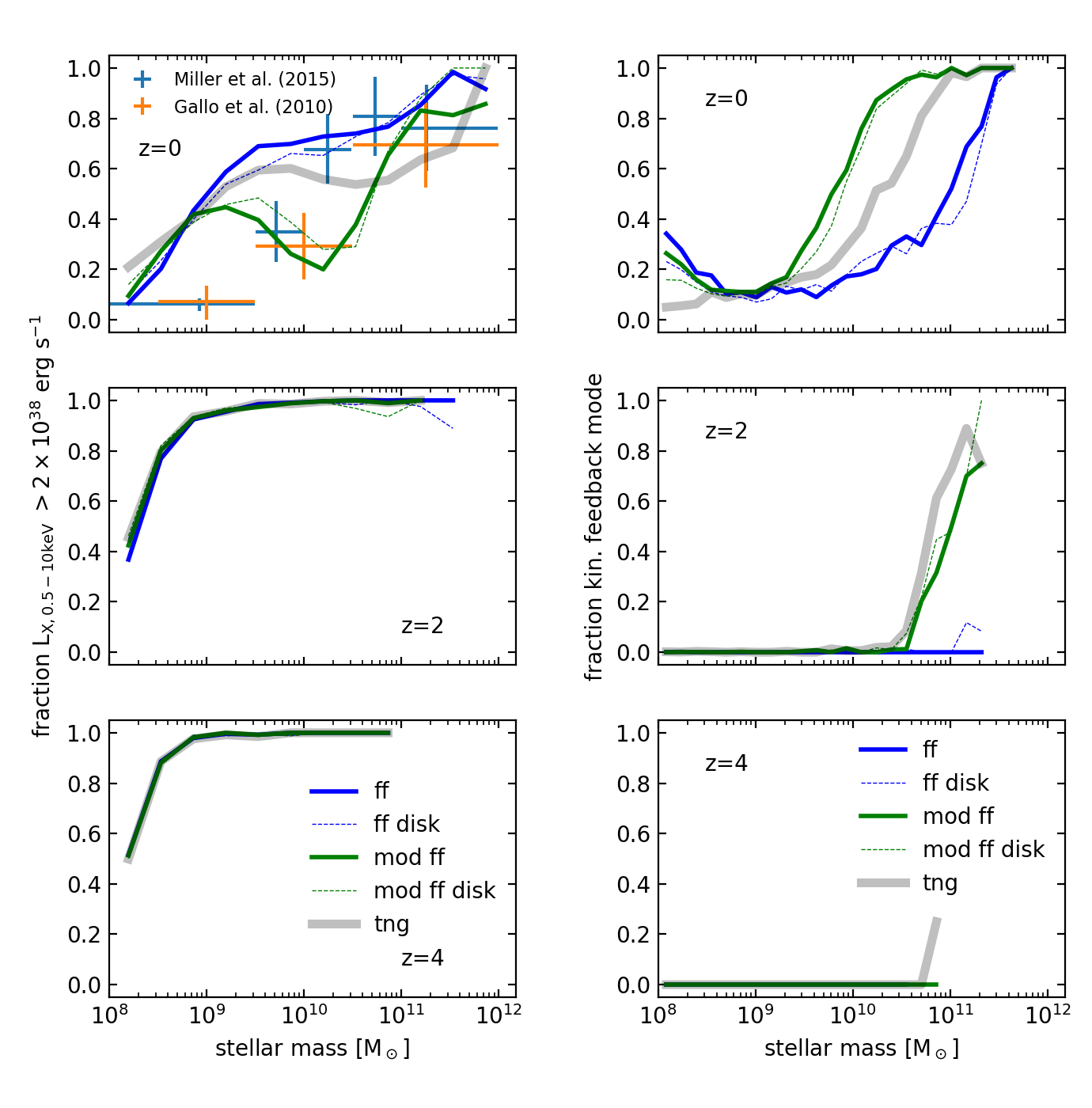}
    \caption{Left:\ Fraction of galaxies with an AGN of X-ray luminosity $>2\times10^{38}\,\mathrm{erg}\,\mathrm{s}^{-1}$ vs stellar mass at redshifts $0$, $2,$ and $4$ (from top to bottom). We compare our results with data from the AMUSE survey \citep{Gallo2010, Miller2015}. Right:\ Fraction of systems with the most massive BH in kinetic feedback mode vs stellar mass at the same redshifts.}
    \label{fig:occupation_fraction}
\end{figure*}

\begin{figure*}
    \includegraphics[]{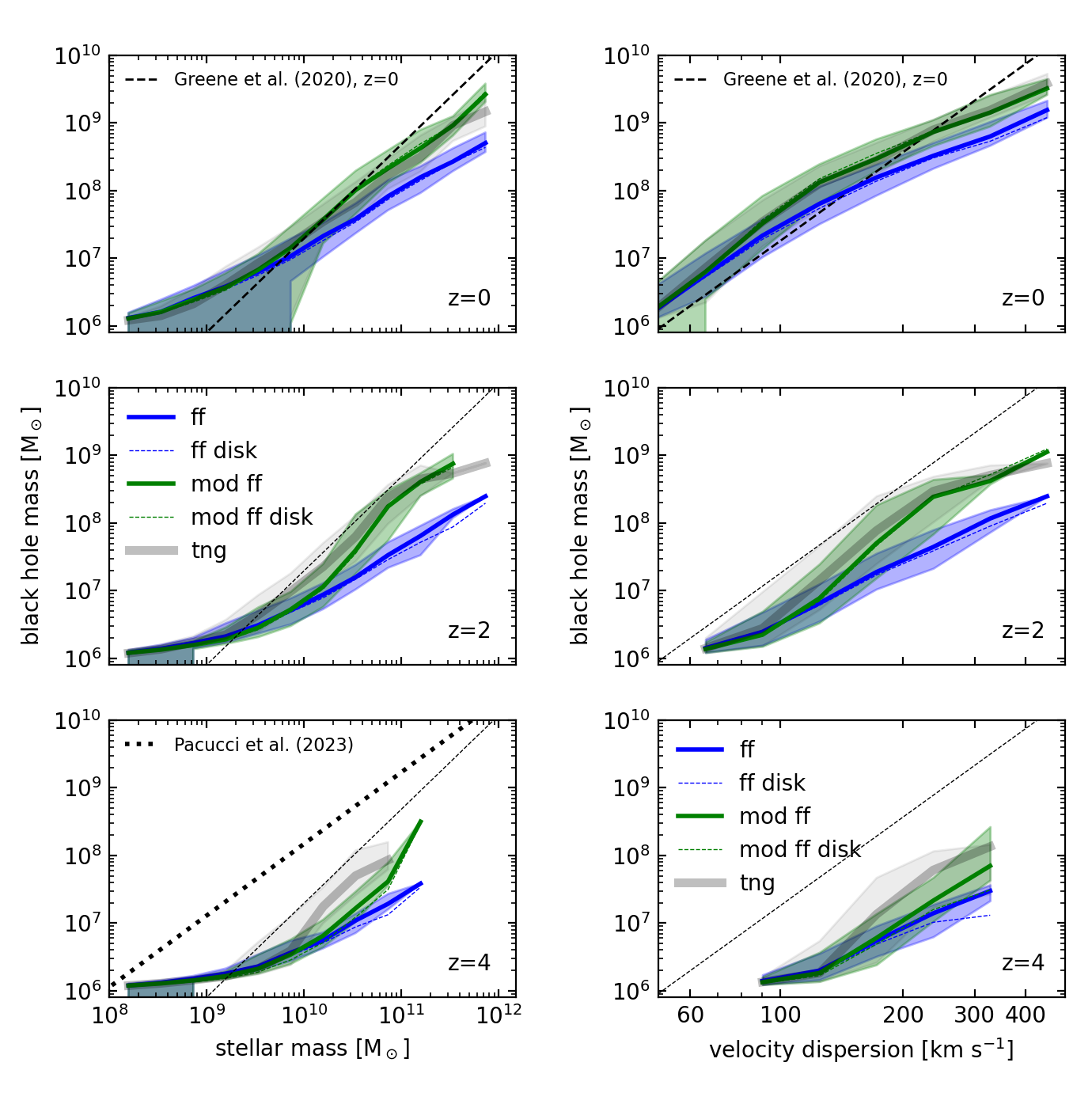}
    \caption{Redshifts of $z=0$ (top), $2$ (middle), $4$ (bottom) BH mass-stellar mass (left) and BH mass-velocity dispersion (right) relations. Note: the \citet{Greene2020} fits to $z=0$ MBHs do not take into account upper limits. We repeat the $z=0$ relation in the higher redshift panels to highlight the redshift evolution in the simulations. The fit from \citet{Pacucci2023} is for $z=4-7$ galaxies.}
    \label{fig:mbh_mstar}
\end{figure*}

One key difference of the different inflow rate estimates is their scaling with BH mass. Since the BH mass function builds up over time, we would therefore expect this mass dependence to translate into a different redshift dependence between the models. Figure~\ref{fig:bhard_z} shows the cosmic BHARD as a function of redshift for the different models. The \texttt{tng} model in grey has the steepest rise at high redshift, but also drops off by about an order of magnitude toward low redshift. The fixed free-fall efficiency model (\texttt{ff}), on the other hand, leads to a shallower rise and relatively small decline of the BHARD at low redshift. We note, however, that the low-redshift decline is steeper for runs of the same model using a higher normalisation, $A$ (not shown here). The \texttt{mod ff} model sits in between the \texttt{tng} and \texttt{ff} models, both in terms of the mass dependence of the accretion rate and in the shape of the corresponding BHARD.

The bottom panel of Fig.~\ref{fig:bhard_z} shows the corresponding BHMD as a function of redshift. The final BHMD results differ only by a factor of about 2 (after calibrating to it), while first differences emerge at $z=4$. At higher redshift, the cosmic mass densities are almost identical, likely facilitated by the use of the same MBH seed model, but also due to very similar BHARDs above redshift $5$. For reference, we show the observationally inferred BHARD integrated up from the \citet{Buchner2015} luminosity functions, assuming a constant radiative efficiency of $0.1$ as well as the mass densities form \citet{Marconi2004, Shankar2004}.  While the redshift evolution is in broad agreement with the \texttt{tng} and \texttt{mod ff} simulations, it is important to note that assumptions of radiative efficiency can substantially shift the data.

Fig.~\ref{fig:mf_lf_xf} shows the distributions of MBH masses, AGN bolometric luminosities, and Eddington ratios (i.e. the accretion rate relative to the Eddington limit, which is the maximum BHs can accrete in our model). The different rows indicate the different redshifts. The mass functions (left column) show an earlier buildup of the high mass end for the \texttt{tng} model at $z=4$ and a flattening between $10^7\,\mathrm{M}_\odot$ and $10^8\,\mathrm{M}_\odot$ that develops later for the \texttt{ff} model. While the $z=0$ mass function matches the fit presented in \citet{Shankar2004} fairly well due to our calibration to the $z=0$ BHMD, the $z=2$ mass functions show an increased number of MBHs compared to \citet{Schulze2015} for the \texttt{tng} and \texttt{mod ff} models. For the bolometric luminosity (middle column) and Eddington ratio (right column) distributions, properties that depend on the instantaneous accretion rate, the changes in the distributions due to a different inflow rate estimate are remarkably small. Hardly any differences arise due to the introduction of the accretion disk model. Overall, the change with redshift is more significant than the difference between models at a given redshift. For reference, we compare the luminosity functions with fits from \citet[][their model A]{Shen2020}. We note, however, that the quasar luminosity function is mostly fit to quasars more rare and more luminous than hosted in our L50 box.

Figure~\ref{fig:occupation_fraction}, left column, shows the fraction of luminous AGN (X-ray luminosity $>2 \times 10^{38}\,\mathrm{erg}\,\mathrm{s}^{-1}$) versus host galaxy stellar mass. There is a general expectation for more massive galaxies to host higher luminosity AGNs, thus the fraction of active galaxies increasing with stellar mass. This is indeed the case with the transition happening between $10^8\,\mathrm{M}_\odot$ and $10^9\,\mathrm{M}_\odot$ at $z=4$ and $z=2$ with only minor differences between the accretion models. For $z=0$, the transition is more gradual, with a substantial fraction of higher mass galaxies not hosting a luminous AGN. The TNG model shows a characteristic plateau in AGN fraction around $10^{10}-10^{11}$~M$_\odot$, similar to previously reported analysis of the IllustrisTNG simulations \citep{Habouzit2019}. Interestingly, the accretion model variations do exhibit a different behaviour with the \texttt{mod ff} model showing a drop at stellar mass of around $10^{10}$~M$_\odot$ and \texttt{ff} a substantially weaker suppression. Compared to observations of \citet{Gallo2010, Miller2015}\footnote{Data retrieved from the analysis script of \citet{Tremmel2024}.}, all our models overpredict AGN in galaxies of stellar mass around $10^9$~M$_\odot$, likely due to overly optimistic seeding in the IllustrisTNG model.

To fully understand this behaviour, it is useful to consider the transition of feedback modes that is kept identical in all models: there is a MBH mass-dependent transition from thermal (high Eddington ratio) to kinetic (low Eddington ratio) feedback mode \citep{Weinberger2017}, which implicitly translates to a relatively sharp transition in stellar mass. Since the kinetic mode is the by far more impactful feedback mode \citep{Weinberger2018} it reduces not only the star formation rate of the host galaxy but also the accretion rate of the MBH. Consequently the bolometric luminosity of these AGN drops, in many cases below our threshold of an active galaxy, thus leading to a drop at precisely the location where most galaxies have their most massive MBH in kinetic mode (right panel of Figure \ref{fig:occupation_fraction}). This effect is also key for the galaxy population (see Appendix~\ref{app:galprop}).

As the \texttt{ff} model has a shallower scaling of the accretion rate with BH mass compared to \texttt{tng}, we expect that, keeping the overall growth and environment the same, low-mass BHs grow more rapidly, while massive BHs grow at a slower pace. Assuming an unaltered stellar mass buildup, this would lead to a shallower slope in MBH-galaxy scaling relations. Fig.~\ref{fig:mbh_mstar} gives the BH mass-stellar mass (left) and BH mass-velocity dispersion (right) relations for the different models, showing median, 10th and 90th percentiles. The dashed line indicates the \citet{Greene2020} fit to MBHs at $z=0$ (their fit excluding upper limits). 
In the $z=0$ relations, the slope of \texttt{ff} is indeed shallower with a power-law slope below unity, while the \texttt{tng} and \texttt{mod ff} models show a close to unity slope above a certain mass range, but still slightly shallower than \citet{Greene2020}. At higher redshift, \texttt{tng} and \texttt{mod ff} show relations below the $z=0$ versions for intermediate masses, while they approach their $z=0$ values at the highest masses. The \texttt{ff} model shows consistently lower mass ratios at higher redshifts. None of the models presented here show over-massive BH mass-stellar mass relations at high redshift \citep{Pacucci2023}.

\subsection{IMBH accretion at high redshift}

In this step of the study, we use the models that best match the global BHMD in our large volume simulation and apply them with the same parameters to study their behaviour at high redshift and for IMBHs. We analysed the high redshift, smaller volume L12.5n768 simulations in which the only other modifications are the seeding parameters: the seeding halo mass is lowered by a factor of $10$ to $M_\mathrm{halo, seed} = 5\times 10^9\,h^{-1}$~M$_\odot$ and the seed BH mass by a factor $1000$ to $M_\mathrm{seed} = 8\times 10^2\,h^{-1}$~M$_\odot$. The choice to lower the seed halo mass below that of the seed mass is meant to avoid over-seeding, as compared to physically motivated seed models \citep[see][for studies using the IllustrisTNG model]{Bhowmick2024}. We note that our main aim in this work is not to study realistic populations of high redshift IMBHs, but only to study the relative differences that different accretion models have on the high redshift IMBH population. 

\begin{figure}
    \includegraphics[]{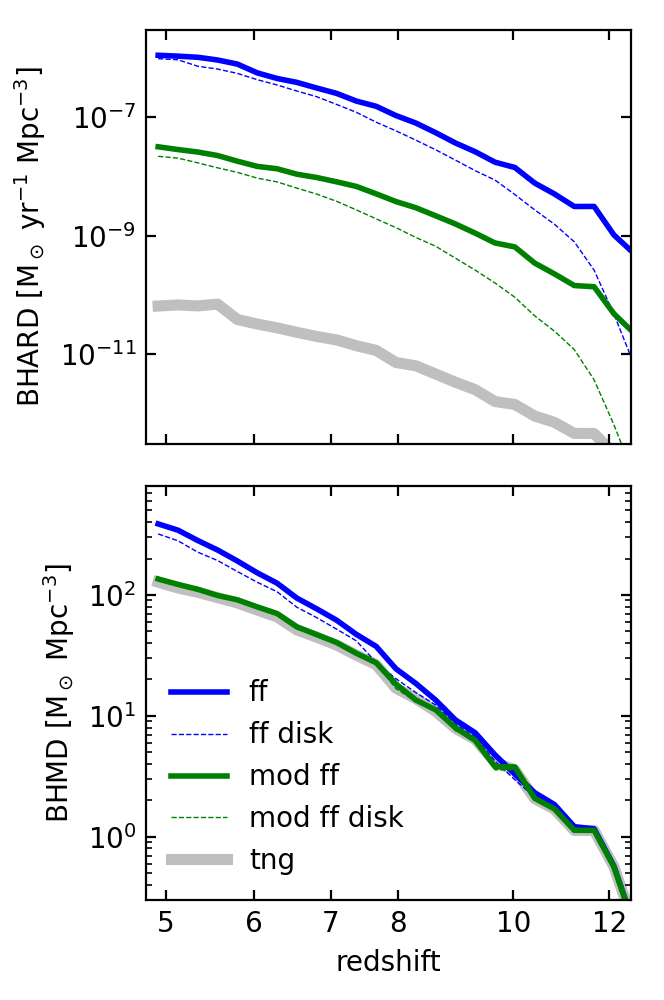}
    \caption{BHARD and BHMD as a function of redshift for the high redshift simulations, L12.5n768. The models that give similar BHARDs in the SMBH regime result in BHARDs that are orders of magnitude different in the IMBH regime. The models with accretion disks show a delay in BHARD compared to the models without accretion disks at the highest redshift, where timescales are shortest.}
    \label{fig:highz_bhard_bhmd}
\end{figure}

\begin{figure*}
    \includegraphics[]{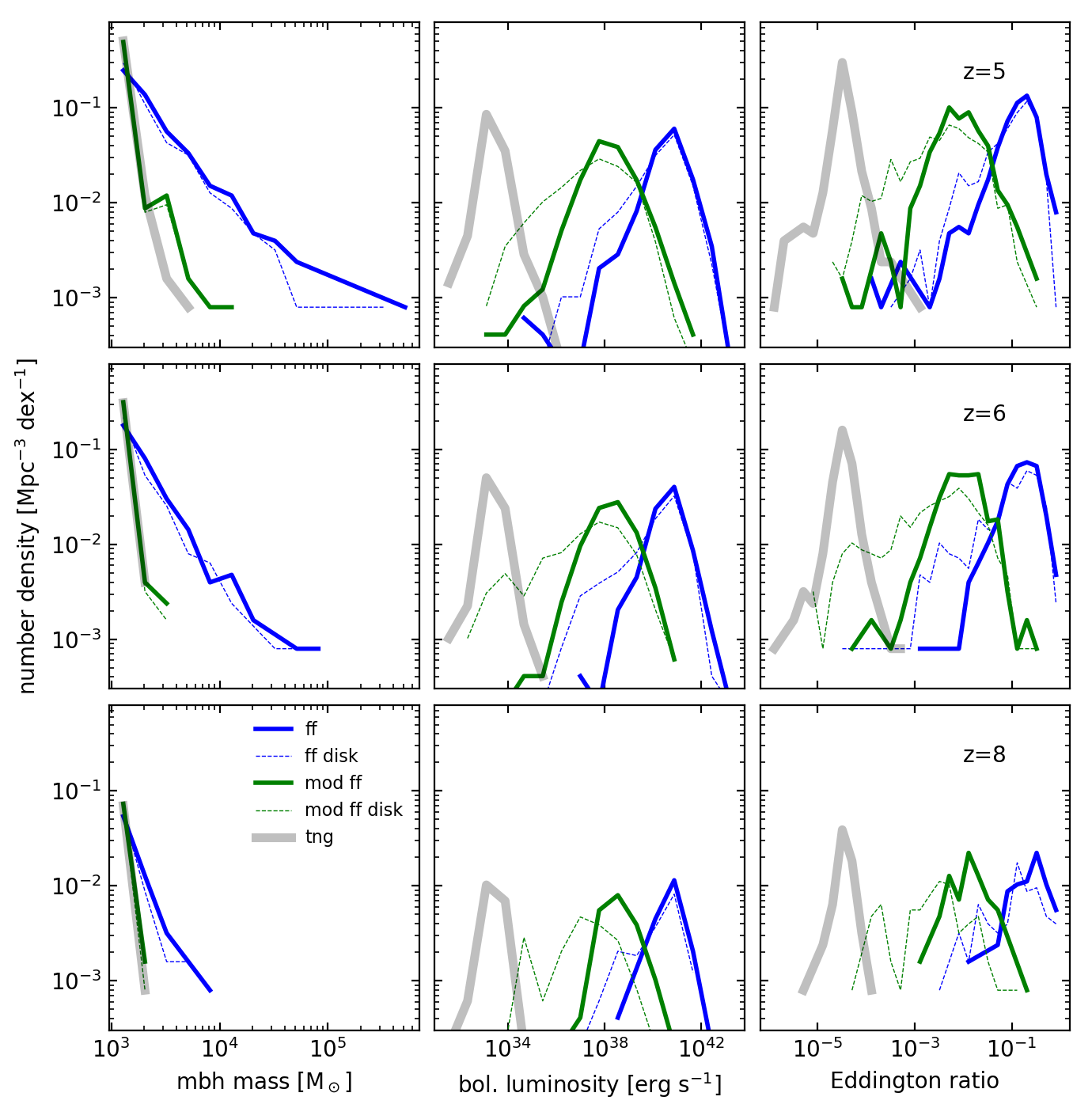}
    \caption{Black hole mass function (left), bolometric luminosity function (centre), and Eddington ratio distribution (right) for IMBHs at redshifts of $5$, $6,$ and $8$ (top to bottom). The dashed lines show the respective models including an accretion disk model. Unlike the SMBH case, the luminosity functions and Eddington ratio distributions differ by orders of magnitude. The addition of an accretion disk model modifies the bolometric luminosity in some regimes. The IMBH mass functions differ mostly due to the growth via accretion in the \texttt{ff} model.}
    \label{fig:highz_mf_lbol_fedd}
\end{figure*}

\begin{figure*}
    \includegraphics[]{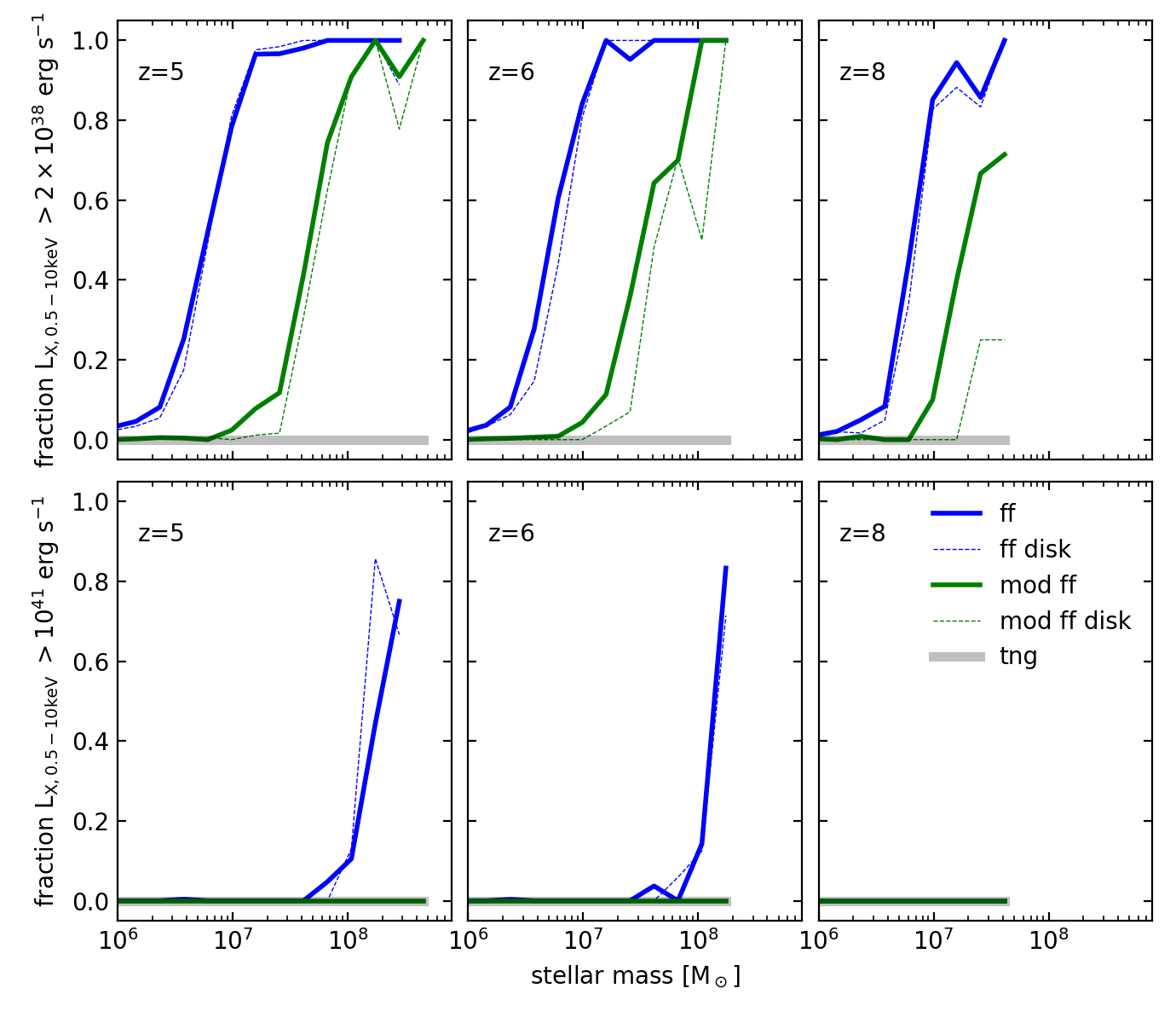}
    \caption{AGN occupation fraction for the IMBH simulations at $z=5$ (left), $z=6$ (centre) and $z=8$ (right). The top panels shows a threshold X-ray luminosity of $2\times10^{38}$~erg~s$^{-1}$ as the comparison in the local Universe. The bottom panel shows the occupation fraction with a threshold of $10^{41}$~erg~s$^{-1}$. While this is is not quite at the luminosity where they would be detectable with future observatories \citep{Marchesi2020}, the trend indicates that AGNs in higher stellar mass systems (not present due to limited simulation volume) might be detectable depending on accretion model.}
    \label{fig:highz_occupation fraction}
\end{figure*}

\begin{figure}
    \includegraphics[]{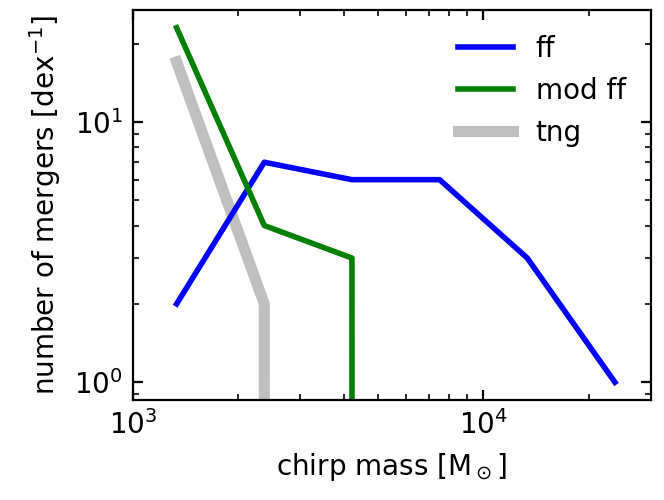}
    \caption{Chirp mass $(M_1M_2)^{3/5}(M_1+M_2)^{-1/5}$ distribution of IMBH mergers in the L12.5 simulations to $z=5$. While the precise normalisation is somewhat arbitrary, the differences due to different accretion rate estimates is significant, with the \texttt{tng} and \texttt{mod ff} models being dominated by seed-mass mergers, while \texttt{ff} is far more distributed due to growth via accretion.}
    \label{fig:merger_rate}
\end{figure}

Figure~\ref{fig:highz_bhard_bhmd} shows the BHARD and BHMD as a function of redshift for the IMBH population. In contrast to the SMBH population shown in Figure~\ref{fig:bhard_z}, the BHARD differs by orders of magnitude depending on the accretion model. This is a direct consequence of the different scaling with BH mass in the different accretion rate estimates. Notably, however, the BHMD is very similar, only gradually diverging starting at $z=8$. This is due to the mass growth in some of these models being dominated by seeding, posing a lower limit. Only the \texttt{ff} model shows growth via gas accretion at similar rates. We note, however, that the frequency of seeding in our model is not well constrained and in its redshift dependence likely incorrect \citep[see][for a comparison between a gas-based and halo based seeding frequency]{Tremmel2017}.

The lack of growth via gas accretion in the \texttt{mod ff} and \texttt{tng} models becomes manifest in the steep mass function shown in the left column of Figure~\ref{fig:highz_mf_lbol_fedd}, with most MBHs being at or close to their seed mass and only minor growth via mergers. The corresponding AGN luminosities and Eddington ratios shown in the middle and right panel of Figure~\ref{fig:highz_mf_lbol_fedd} are relatively low, however, differing by several orders of magnitude between \texttt{tng} and \texttt{mod ff}. In the \texttt{ff} model, the Eddington ratio peaks at values larger than $0.1$, and corresponding luminosities close to their Eddington luminosity.

For direct comparison to the SMBH case, we first adopt the same threshold of an active galaxy with X-ray luminosity of $2\times10^{38}$~erg~s$^{-1}$ for an occupation fraction analysis. Since this is clearly not detectable at high redshift, we also show the occupation fraction with a threshold of $10^{41}$~erg~s$^{-1}$ in Figure~\ref{fig:highz_occupation fraction} as a function of stellar mass. Note that unlike the SMBH case, all of the AGNs in these systems are in thermal feedback mode, thus specific imprints due to feedback mode switches are not expected. The transition from no AGN to all AGN occurs at strikingly different stellar masses for the different accretion models.

In Figure~\ref{fig:highz_mf_lbol_fedd} we have shown that the IMBH mass function is quite different for the different models. It is therefore a natural question to ask if gas accretion can also alter the merger rate of two IMBHs with given mass. In Figure~\ref{fig:merger_rate}, we show the chirp mass,
\begin{align}
\mathcal{M} = \frac{(M_1M_2)^{3/5}}{(M_1+M_2)^{1/5}},
\end{align}
of IMBH mergers until $z=5$ for the three models (since the accretion disks make practically no difference in the mass functions, we do not show them here). It should be noted that, due to the chosen seeding parameters, the number of mergers is relatively limited (of order 30 per simulation). Furthermore, we operated with an instantaneous merger model, not taking dynamical friction on spatially resolved scales nor the complex inspiral process on unresolved scales and its delays into account \citep[see][for a more thorough modelling]{Kelley2017}. Thus, the result should not be interpreted as an absolute prediction, but only the relative changes due to accretion model variation are relevant. Indeed, the models with a steep mass function \texttt{tng} and \texttt{mod ff} have a merger distribution strongly peaked at the seed mass, with few mergers of higher mass BHs, while the \texttt{ff} model, with its considerable mass growth via accretion has a broader chirp mass distribution.

\section{Discussion}
\label{sec:discussion}

Here, we study the effects of different accretion models in cosmological simulations for both SMBH and IMBH gas accretion. While the models are inspired by high-resolution simulations and analytic considerations, they are intentionally built to be able to parametrise different scalings and represent different mechanisms driving the gas inflow. For the present work, we adjust the parameters to roughly match the SMBH mass density at low redshift (Fig. \ref{fig:bhard_z}). We analyse the remaining differences in SMBH growth over cosmic time and explore how these parameters would affect the growth of IMBHs at high redshift.

The pre-factor in the \texttt{ff} model, $A = 0.001$ is somewhat smaller than in other works using similar models \citep{Ehlert2023, Wellons2023}, however, we note our larger accretion region compared to cosmological zoom simulations and lack of other criteria such as only accreting sufficiently cold gas, might explain the need for a lower efficiency. In the \texttt{mod ff} model, $A=100$ should be interpreted as the radius dependent mass inflow rate slope $\alpha = -0.5$ not reaching all the way to the Schwarzschild radius, but transitioning to a constant inflow rate with radius at $r = A^2\, r_s  = 10^4\, r_s$. Note that since $r_s \propto M$, the scaling of the inflow rate with BH mass changes as $\dot{M}\propto M^{0.5-\alpha}$, driving the differences between \texttt{mod ff} ($\dot{M}\propto M$) and \texttt{ff} ($\dot{M} \propto \sqrt{M}$). The Bondi accretion rate scales as $\dot{M}\propto M^2$ and is additionally modified by a sound speed dependence and a different way the density is estimated, making the direct comparison more difficult. 

\subsection{Non-uniqueness of the MBH gas accretion in a galaxy formation context}

In the SMBH regime, the  substantial changes made to the accretion model have only moderate impacts on the MBHs over cosmic time once the normalisation has been set to roughly match the BHMD at $z=0$. While the qualitative shape of the BHARD is largely dictated by the gas accretion rate onto the galaxy \citep{vandeVoort2011}, the precise shape changes with mass scaling (Fig.~\ref{fig:bhard_z}), in particular the redshift of peak AGN activity and the magnitude of the reduction to $z=0$. 

The bolometric luminosity and Eddington ratio distribution (Fig.~\ref{fig:mf_lf_xf}) are remarkably insensitive to changes in the accretion rate estimate. Correspondingly, the occupation fraction of luminous AGN (Fig.~\ref{fig:occupation_fraction}) is also insensitive to a change of accretion models, with the exception of a feedback related transition from thermal to kinetic feedback in our model. We note, however, that this transition needs to be readjusted when calibrating to realistic galaxies (see Appendix~\ref{app:galprop}), which will likely reduce this difference. 

This behaviour indicates that the MBH growth ends up becoming self-regulated by feedback in many regimes. The local conditions then adjust until a certain balance between liberated feedback energy and large-scale inflow rate can be reached, which effectively sets a certain accretion rate if the feedback model is fixed. This resembles situations, for example, in star formation, where the total stellar mass of a galaxy is relatively independent of the local star formation efficiency assumed for the ISM model - instead it is determined by the feedback and the cooling rate out of the CGM.
As a consequence, luminous AGN at a given redshift are far more sensitive to the conditions for efficient AGN feedback than to accretion rate scalings. Based on these findings, it seems difficult to distinguish different accretion models based on luminous AGNs powered by SMBHs. 

The BH mass functions at $z=0$ are relatively similar. At $z=2$ and $z=4$, differences are present at the high mass end (Fig.~\ref{fig:mf_lf_xf}, left column). The dominance of mergers for the mass growth of high mass SMBHs at low redshift \citep{Weinberger2018} likely erases some of these differences by $z=0$. This is, of course, to a certain degree a consequence of the calibration of the model against the $z=0$ BHMD; namely, the integrated mass function. Consequently, the BH mass-stellar mass and BH mass-velocity dispersion relations (Fig.~\ref{fig:mbh_mstar}) differ only mildly with slightly flatter relations for the \texttt{ff} model. Yet, given large observational uncertainties and biases, it is hard to imagine how they would be suited to reveal the nature of gas inflow. In this context, it is important to keep in mind that we keep all other parameters constant. A shallower mass dependence, when normalised to a reasonable $z=0$ BHMD, does not produce over-massive BHs relative to the $z=0$ relations at any redshift. Thus, with the model variations shown here, we are unable to reproduce the scaling inferred by \citet{Pacucci2023}, nor provide a sufficient scatter to explain the reported detections \citep{Li2024}. Redshift dependent scalings e.g. in the radiative efficiency or a more efficient quasar mode feedback that more efficiently regulates accretion after some MBHs have grown overmassive might be needed to simultaneously produce over-massive high-redshift BHs at $z=4-7,$ while not overgrowing them toward $z=0$.

The similarity in MBH and AGN properties between different accretion models does, however, not imply that the choice of accretion model is irrelevant in future cosmological simulations. Most importantly, the MBH seed model was kept constant throughout this study. We used the IllustrisTNG model, which by construction completely omits the IMBH regime by placing a new MBH with mass $\sim10^6$~M$_\odot$ in halos above a given mass. Potential growth starting from a more physically motivated formation scenario based on gas and radiation properties, however, will likely lead to an accretion model dependent growth to $10^6$~M$_\odot$, resulting in different occupation fractions of SMBHs more massive than $10^6$~M$_\odot$. This, in turn, will cause additional differences in the subsequent evolution of the simulated SMBHs. Future simulations that aim at predicting the MBH population from low mass seeding scenarios will have to take this into account.

\subsection{IMBH gas accretion}
Unlike in the SMBH regime to which the parameters were calibrated, the different models cause massively different outcomes assuming $10^3$~M$_\odot$ seeds. The resulting BHARD changes by orders of magnitude for the different models (Fig.~\ref{fig:highz_bhard_bhmd}). The resulting bolometric luminosity and Eddington ratio distributions are orders of magnitude apart (Fig.~\ref{fig:highz_mf_lbol_fedd}), leading also to substantially different occupation fractions provided the threshold luminosity is low enough (Fig.~\ref{fig:highz_occupation fraction}). 
The employed X-ray luminosity thresholds are lower than expected detection limits with Athena and AXIS. Future surveys with these telescopes can detect AGNs with soft ($0.5-2$~keV) X-ray luminosities down to $\sim10^{43}$~erg~s$^{-1}$ and $\sim10^{42}$~erg~s$^{-1}$ at $z\sim 5$, respectively \citep{Marchesi2020}. If we used thresholds more in line with (future) X-ray detectable AGNs, the occupation fraction in our $L12.5$ boxes would be vanishing for all models. However, we note that this is merely an artifact of the small volume, and consequently low-mass host galaxies simulated here. Simulating larger volumes would naturally produce more massive host galaxies as well, with slightly more massive IMBHs (of mass $10^5-10^6$~M$_\odot$ at $z=5$). Since the Eddington ratio scales differently with BH mass for the different models ($M^{-1/2}$ for \texttt{ff}, $M^0$ for \texttt{mod ff} and $M^1$ for \texttt{tng}), the differences in Eddington ratio distribution would be less pronounced for more massive IMBHs (assuming no other effects). However, the differences might still be present, producing a sweet-spot between detectability and differences in accretion rate in this mass range. We thus argue that future studies of the detectability of AGNs powered by IMBHs, not just in X-ray but also in optical wavelengths \citep[e.g.][]{Cann2019} need to carefully take uncertainties in the accretion rate estimate into account.

We note that our simulations are not designed to model gas inflows onto individual IMBHs in great detail, but rather to explore the consequences of different assumptions on a larger population of IMBHs. Dependency on the IMBH formula, however, highlights the importance of dedicated studies of high redshift gas accretion onto IMBH. A dependence on inflow model implies that resolved scales do not include the bottleneck of gas accretion, and smaller scales play a crucial role. With sufficient resolution to capture these scales, however, it is expected that the accretion rate converges independently of accretion formula. \citet{Gordon2024}, for example, find a converged accretion independent of accretion scheme for $100$~M$_\odot$ IMBHs with simulations reaching resolutions of around $10^{-3}$~pc, i.e., at significantly higher resolution than achievable with uniform volume approaches.

Interestingly, the dramatic change in accretion rate does not in all cases carry over to the BHMD, indicating that its buildup has a floor set by seed formation. For the more optimistic accretion scenarios, however, the mass growth via gas accretion of individual BHs can be substantial, flattening the BH mass function (Fig.~\ref{fig:highz_mf_lbol_fedd}, left column). This accretion driven growth also modifies the frequency of IMBH mergers of given chirp mass (Fig.~\ref{fig:merger_rate}), highlighting the importance to take it into consideration when interpreting merger rates, for instance, from future LISA gravitational wave detections \citep{Sesana2007, Amaro-Seoane2017}. In addition, uncertainties in the binary inspiral phases \citep{Kelley2017, Siwek2020} need to be considered, which we neglected in this work due to the instant merger assumption for MBH binaries. 

However, for reliable predictions of the growth history, dynamical friction and BH recoil kicks are two aspects that are missing in the employed model, which are also of particular importance for IMBHs. Since dynamical friction is relatively inefficient at low masses, IMBHs are not necessarily residing in the centre of galaxies but rather in lower density outskirts, leading to lower accretion rates and consequently mass growth. Also,  growth via mergers can be impeded via recoil kicks, which can lead to merged IMBHs to escape the host halo. We leave the study of these effects to future work.

It is important to note here that the presented seeding is just a simple change in the parameters of the IllustrisTNG seeding prescription and not a prediction for any specific pathway \citep[see][for a review]{Volonteri2010}. However, it provides a reasonable environment to study the potential for growth of IMBHs via accretion. Considering the BHMD in our simulation, accretion only represents a fraction of the seeded BHs, and a more carefully considered IMBH seeding prescription is needed to fully assess the fractional growth via accretion on these systems along with previously discussed observational constraints. Improved MBH formation models should accurately predict seed locations, redshifts \citep{Bellovary2011}, frequency \citep{Habouzit2016} and seed mass function \citep[e.g. as modelled in][]{Ni2022}.

\subsection{The role of accretion disks}

Apart from varying the accretion model, we also ran the different inflow models with an additional model for unresolved accretion disks. Its main function is as a gas reservoir that is filled by the gas inflow and drained over a given timescale. The natural expectation of such a model is that fluctuations in the instantaneous inflow rate will be smoothed out, while the overall accretion rate should not be affected since it is limited by the inflow rate, not the accretion disk (with the exception of the Eddington-limited regime). Indeed for the mass growth, we see very little difference between the mass growth with and without an accretion disk model, only some initial delay due to the accretion disk timescale at the highest redshifts. In quantities proportional to the instantaneous accretion rate, i.e. luminosity and Eddington ratio, no substantial modifications arise in the SMBH regime. In the high redshift IMBH regimes, larger differences are visible. We speculate that this is more related to the higher redshifts and correspondingly shorter structure formation timescales (becoming comparable to the accretion disk timescale) rather than to the mass of the MBHs. 

It should be noted that the implemented accretion disk model is very simple and lacks some of the features that might systematically alter the AGN population and the MBH growth behaviour. In particular variations to the radiative efficiency have the potential to have systematic effects not covered here. These variations could be caused either by varying MBH spin and thus a change in innermost stable circular orbit (ISCO) or by modifications in the accretion disk, e.g., in a slim disk accreting at super-Eddington rates. We leave explorations in this direction to future work.

\section{Summary and conclusions}
\label{sec:conclusions}

We present a new, simple model for MBH accretion. We compare its performance and results to those of the IllustrisTNG model. The key aspects of our model are:

\begin{enumerate}
    \item The inflow estimate is based on a free-fall time, which scales more weakly with BH mass than the Bondi formula. It also scales linearly with an enclosed gas mass.
    \item The surrounding gas properties are estimated from gas within a fixed proper distance from the BH.
    \item The accretion rate is integrated as a source term for each individual cell, in line with source and sink terms in the finite-volume approach.
    \item A model for a radial dependence of the unresolved part of the mass inflow rate is also introduced. Its scaling affects the overall BH mass scaling.
    \item A simple sub-grid accretion disk model is implemented.
\end{enumerate}

The normalisation parameter in these models was chosen to result in a roughly comparable $z=0$ BHMD. Fixing this parameter, the key effects in the SMBH population are:

\begin{enumerate}
    \item At fixed final BHMD, the BHARD can vary only moderately. Interestingly, the differences at $z>5$ are very small when keeping seeding and AGN feedback the same. The BHARD at cosmic noon versus at $z=0$ can vary somewhat, depending on accretion rate estimate, with a less pronounced peak and less dimming toward $z=0$ for less efficient accretion (in the free fall-based model in our case).
    \item The quasar luminosity functions are remarkably similar for the different accretion models. The main effect, most obvious in the occupation fractions at $z=0$, is a non-linear interaction with AGN feedback. The transition to efficient AGN feedback, which differs for the different accretion models, has a more substantial effect on luminous AGNs than the inflow rate estimate (within the bounds tested in this work). In future works, a recalibration of the AGN feedback will be needed to match galaxy population constraints to properly assess the uncertainties in the MBH population. Notably, even prior to AGN feedback becoming a relevant factor for the galactic star formation rate, our results indicate an AGN feedback induced self-regulation of the accretion rate.
    \item The BH mass-stellar mass and BH mass-velocity dispersion relations change moderately, with a slight flattening of both relations for the free-fall based model. Notably, provided the overall normalisation is calibrated to the $z=0$ mass density, a weaker mass dependence of the accretion rate does not lead to substantially over-massive BHs at the low-mass end of the BH mass-stellar mass relation. This is a trend suggested by observations  at high redshift \citep{Pacucci2023}. We speculate that a variable radiative or feedback efficiency would be required to achieve this, while making sure not to violate the constraints on the X-ray background.
\end{enumerate}

We therefore conclude that the precise accretion rate formula used in galaxy formation simulations  only has minor impacts on the galaxy and the SMBH populations once it is tuned to match the BHMD at $z=0$.
However, when we apply the different accretion rate estimates to a high-redshift simulation modelling an IMBH population with a seed mass around $10^3\,\mathrm{M}_\odot$, drastic changes are evident. In particular, we find:

\begin{enumerate}
    \item Accretion rate densities differ by four orders of magnitude, while the mass growth is less affected due to growth via seeding. This indicates that our lack of understanding of the accretion mechanism leads to substantial uncertainties for possible signatures and growth histories of high redshift IMBHs.
    \item Mass functions of IMBHs change through gas accretion for free-fall based accretion. This has important implications for IMBH merger rates in different mass bins and consequently for the interpretation of future LISA detections of IMBH mergers \citep{Sesana2007}.
    \item AGN luminosities differ by orders of magnitude depending on the model. We only show occupation fractions with luminosity thresholds below the detection limit of future X-ray surveys with Athena or AXIS \citep{Marchesi2020}. This is due to our relatively small simulation volume of $12.5\,h^{-1}$~Mpc side length that only contains low-mass IMBHs at this redshift. However, the trend indicates that slightly more massive BHs that would be present in larger volumes would be detectable at $z\sim 5$ depending on accretion rate estimate. More works that include simulations of larger volumes are needed to explore whether these AGNs could constrain the inflow rate scaling onto MBHs. Such studies would also help us explore  how sensitive these object are to variations in the quasar mode feedback.
\end{enumerate}

These changes highlight the need to understand accretion, in particular, the BH mass dependence of the gas inflow to accurately model the high redshift IMBH regime. While we kept the seeding model fixed for the different SMBH simulations in this work, future simulations that model the cosmic evolution of MBHs from light seeds are also likely to show substantial accretion model dependent growth in SMBHs caused by the modified early growth history. Further studies combining this effort with effective seed models \citep[e.g.][]{Bhowmick2024} are clearly needed to explore the possible IMBH evolution scenarios at high redshift and their consequences for luminous AGN and IMBH merger events, as well as the MBH population in the Local Universe.
Further  aspects missing from the employed model include: the  treatment of (resolved) MBH dynamics, allowing for off-centre BHs, as well as recoil kicks from merging MBH binaries, which are expected to moderate the growth from accretion and mergers, respectively.

Overall, this new model represents a general gas inflow formula that parametrises different dependences on BH mass and different normalisations. It was developed to match the sprit of the `Learning the Universe' project; specifically, it is meant to be applied as an effective model with parameters to be constrained by data or marginalised over in order to infer galaxy formation and cosmological parameters from structure formation. We deliberately chose to only use reliable gas properties in the simulation that are only weakly dependent on other model choices, such as in the employed ISM modelling \citep[see][for issues with the Bondi estimate]{Sivasankaran2022}. Measuring the simulation quantities at a fixed aperture also allows the model to be tested using high-resolution inflow simulations \citep{Guo2023, Guo2024}.

\begin{acknowledgements}
We thank the anonymous referee for the suggestions that helped to improve the manuscript.
RW acknowledges funding of a Leibniz Junior Research Group (project number J131/2022). 
This work was supported by the Simons Collaboration “Learning the Universe”.
R.W. acknowledges support from the NSF via XSEDE allocation PHY210011 in the early phases of this project.
L.B. acknowledges support from NASA award \#80NSSC22K0808 and NSF award AST-2307171.
G.L.B. acknowledges support from the NSF (AST-2108470 and AST-2307419, ACCESS), a NASA TCAN award, and the Simons Foundation through the Learning the Universe Collaboration. 
\end{acknowledgements}


\begin{appendix} 

\onecolumn
\section{Impact on the galaxy formation model}
\label{app:galprop}

\begin{figure*}[h!]
\includegraphics{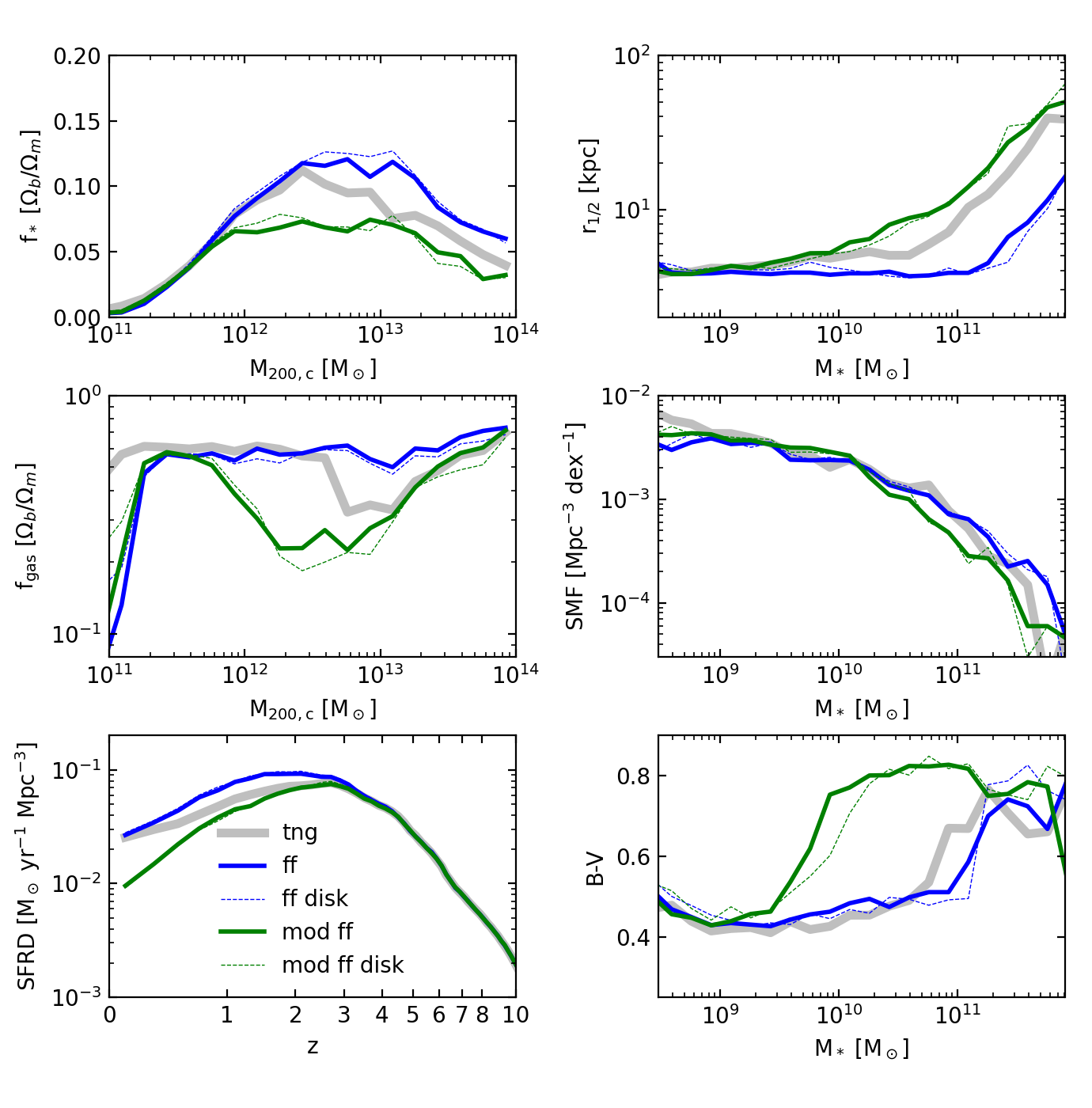}
\caption{Galaxy properties for the different models: stellar mass fraction vs halo mass ($M_{200,c}$); stellar half mass radius vs stellar mass, gas mass fraction within $R_{500,c}$ vs halo mass; stellar mass function; star formation rate density and B-V colours as a function of stellar mass for the L50n768 simulations with different accretion models. The stellar mass and colours are measured within twice the stellar half mass radius; except for the SFRD, all relations are shown at $z=0$.}
\label{fig:galprop}
\end{figure*}

As MBHs have a substantial impact on the evolution of their host galaxies \citep{Fabian2012}, it is reasonable to assume that changing the gas accretion formula has an impact on the galaxy population. We explore this using the calibration plots in IllustrisTNG \citep{Pillepich2018}, and show the median trends of stellar mass fraction versus halo mass, gas fraction versus halo mass, star formation rate density versus redshift, stellar half mass radius versus stellar mass, stellar mass function and B-V colour versus stellar mass for the different models in Figure~\ref{fig:galprop}. While there are some notable differences in all of these plots, we argue that they originate from the same single cause: the precise mass scale where efficient AGN feedback sets in.

The IllustrisTNG AGN feedback model, which is used in all of the presented model variations, has a two-mode AGN feedback, where the dividing line between the relatively inefficient thermal feedback and the efficient kinetic feedback is a mass-dependent Eddington ratio threshold $\chi$ given by 

\begin{align}
    \chi = \min\left[ 0.002 \left(\frac{M}{10^8 \,\mathrm{M}_\odot} \right)^2, 0.1\right],
\end{align}
where $10^8\,\mathrm{M}_\odot$ denotes a characteristic BH mass scale \citep{Weinberger2017}. This leads to a clear transition with less massive BHs predominantly powering inefficient thermal feedback and more massive BHs predominantly powering efficient kinetic feedback \citep{Weinberger2018}. Consequently, the stellar and halo mass scale at which AGN feedback becomes efficient depends strongly on the mass of the BHs in them, and most of the differences in the resulting scaling relation can be tied back to the changed BH mass-stellar mass and BH mass-halo mass relations.

Considering the BH mass-stellar mass relation of Figure~\ref{fig:mbh_mstar} in more detail, the BHs reach the mass of $10^8\,\textrm{M}_\odot$ at higher halo and stellar mass in the \texttt{ff} model, and at lower masses in the \texttt{mod ff} model (compared to the \texttt{tng} model). This leads to massive galaxies in the \texttt{ff} model remaining star forming up to higher stellar and halo masses, leading to higher stellar mass fraction in the high-mass end (top left panel of Figure~\ref{fig:galprop}). The lack of feedback implies the retention of higher gas fractions at halo masses around $10^{13}\,\mathrm{M}_\odot$, and the additional star formation in these systems shows up as slightly higher star formation rate densities at low redshift. The later shutoff of star formation in massive galaxies leads to fewer dry mergers that could increase stellar half mass radii in high mass galaxies and consequently more compact galaxies in this regime, as well as a transition from red to blue colours happening at higher masses. For the galaxies in the \texttt{mod ff} model, the efficient AGN feedback sets in at lower halo masses, as clearly visible in the drop in gas fractions above $10^{12}\,\mathrm{M}_\odot$. This leads to a slight drop in star formation rate density toward redshift 0, as well as a transition to red, quiescent galaxies at a lower stellar mass. It leads to a slight reduction in stellar mass fraction and increase in stellar half mass radii, likely due to the increased dry-merger rate dominating the stellar mass evolution in this regime over in-situ star formation. Thus, for both inflow rate variations, simply adjusting the transition mass would likely compensate for the effects on the galaxy population. We note there are some other minor changes in the low-mass population, however, we refrain from over-interpreting these due to the moderate resolution simulations used in this work. Finally, the unresolved accretion disk models have a negligible effect on the galaxy population, similar to their role in the global BHMD buildup over cosmic time.

In summary, changing the accretion model while keeping all other model choices fixed does result in considerable changes in galaxy scaling relations. However, it is possible to come up with modifications in other model choices and parameters that would largely compensate for the inflicted effects on the galaxy population. The parameters in this case can be considered free parameters that are hard to interpret physically, highlighting degeneracies in model choices in galaxy formation models. Ultimately, individually testable parametrisations could alleviate this problem in future models. 

\FloatBarrier

\section{The environment of MBHs}

\begin{figure}[h!]
    \centering
    \includegraphics[]{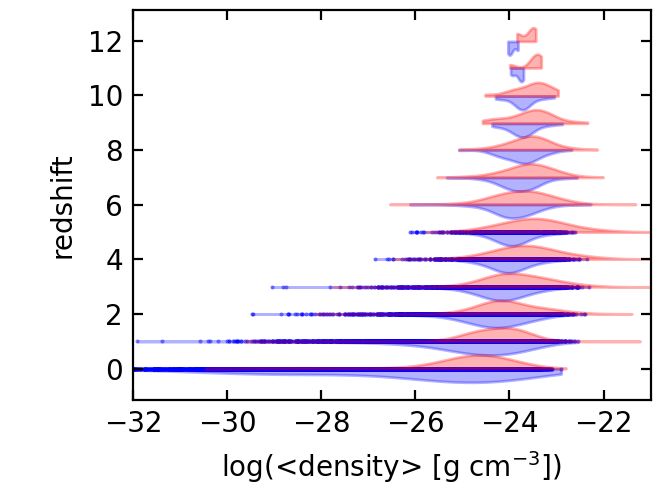}
    \caption{Environment of MBHs at different redshifts. Density is the volume averaged density over the accretion region. Blue: \texttt{ff} with fixed accretion radius; red: \texttt{tng} model with fixed weighted number of neighbours.}
    \label{fig:environment}
\end{figure}

One of the goals of choosing a fixed aperture to measure the gas properties surrounding the MBH to estimate the accretion rate is to provide a well-defined problem that can be replicated at higher resolution. These higher resolution simulations can then be used to test the validity of the employed model, and to inform future model refinements. This way, cosmological and isolated studies can be combined in a more seamless way to understand the impact of physical mechanisms on astrophysical observables. Figure~\ref{fig:environment} shows the aperture-averaged density around MBHs for different redshifts. Interestingly, the density at high redshift is remarkably constant up to redshift 4, with densities getting gradually lower toward low redshift, as well as the distribution function becoming broader. 

Changing from a fixed weighted number of neighbours to a fixed proper distance from the MBH has consequences for the density estimate. We show the volume averaged densities in the \texttt{tng} model run in Figure~\ref{fig:environment}, for comparison. At the highest redshifts, the density distributions are slightly overestimated compared to the fixed aperture runs. This is due to the averaging length becoming systematically smaller, thus probing a smaller region (see Figure~\ref{fig:schematic}). Practically this means that choices of aperture has only impacts to accurately model the highest redshift gas accretion of MBHs. It is important to note, however, that at these high redshifts where the density estimate differs, the mass growth is overall subdominant. Nonetheless, these differences might change the predicted luminosities of high redshift AGNs. On the opposite end, at low redshift for the least luminous AGN, there are differences as well, with the fixed-aperture estimate extending to far lower densities. This is likely due to feedback clearing the surroundings, and while the neighbour search automatically readjusts its search radius up to larger radii, the fixed-aperture estimate simply capture the low-density centre. This, however, will only affect the precise luminosities of low-luminosity or dormant AGN at low redshift, and not affect the growth of SMBHs or their host galaxies. 

\FloatBarrier

\section{Numerical convergence} 

\begin{figure*}[h!]
\includegraphics{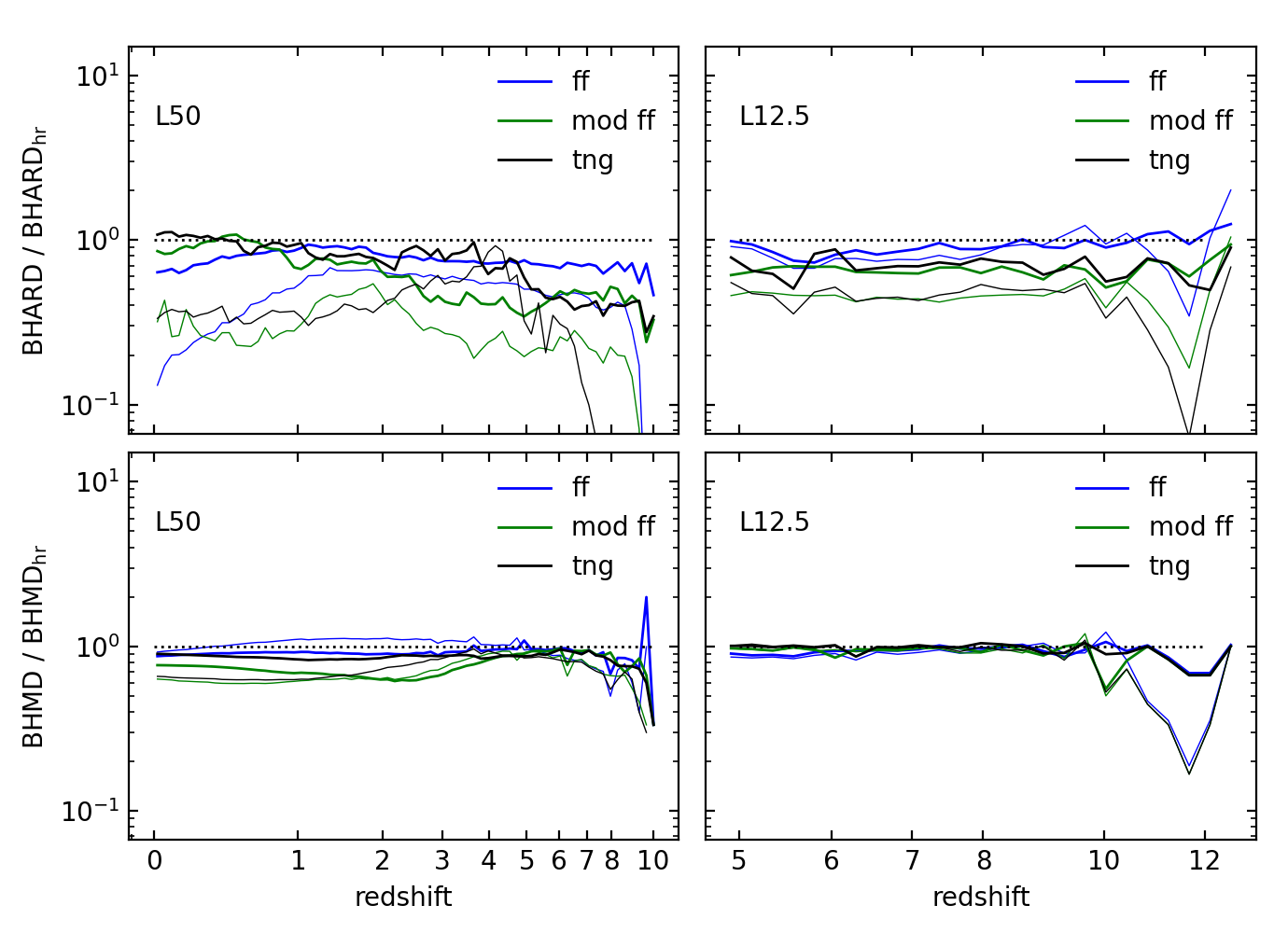}
\caption{Change in BHARD and BHMD relative to the highest resolution simulations for the low redshift (left) and high redshift (right) runs.}
\label{fig:convergence1}
\end{figure*}

To study the numerical convergence of the new methods, we performed each simulation at $3$ resolution levels (see Table~\ref{tab:ics}). We show the BHARD  and BHMD  relative to the highest resolution run in Figure~\ref{fig:convergence1} for the L50, SMBH runs and for the L12.5, IMBH runs in the left and right panels, respectively. While the accretion rates mostly increase with higher resolution, the resolution effects are relatively moderate, except for the lowest resolution L50 runs.

\FloatBarrier

\onecolumn
\section{Environment of IMBHs}
\begin{figure*}[h!]
    \centering
    \includegraphics[width=1.0\linewidth]{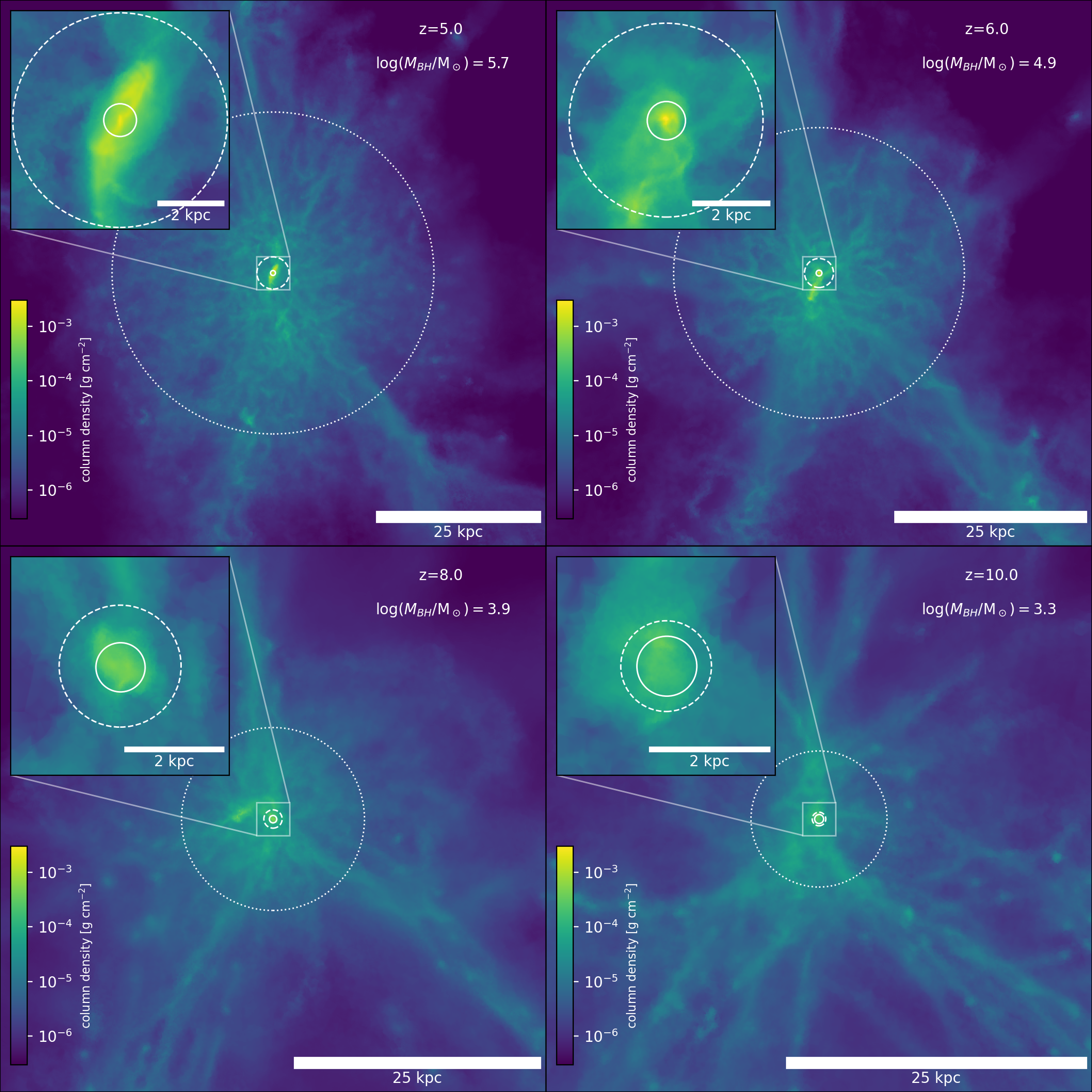}
    \caption{Column density projection of halo~0 at redshift 5 and its progenitors at $z=6,8$ and $10$ of the L12.5n768 \texttt{ff} simulation. The dotted circle indicates $R_{200,c}$, the dashed circle $0.1\,R_{200,c}$, the solid circle the accretion region of the closest BH particle. The scalebar denotes proper distance.}
    \label{fig:gas_projecction}
\end{figure*}
To illustrate the environment of IMBHs at high redshift, Figure~\ref{fig:gas_projecction} shows the gas column density in halo~0 at redshift 5 and its main progenitor at redshift 6, 8 and 10. The dotted, dashed and solid circle show $R_{200,c}$, $0.1\,R_{200,c}$ and the accretion radius, respectively. While the accretion radius at $z=10$ approaches the galaxy radius, the gas structure at these early times is clearly limited by the mass resolution and modelling limitations.

\FloatBarrier
\twocolumn

\end{appendix}

\end{document}